\documentclass[aps,onecolumn,prd,showpacs,nofootinbib]{revtex4}
\usepackage{amsmath}
\usepackage{graphicx}
\usepackage{subfigure}
\usepackage{dcolumn}
\usepackage{bm}
\usepackage{amssymb}
\usepackage{latexsym}

\newcommand{\ie}{\emph{i.e.~}}

\newcommand{\mP}{m_{\mathrm{Pl}}}

\def\spose#1{\hbox to 0pt{#1\hss}}
\def\lta{\mathrel{\spose{\lower 3pt\hbox{$\mathchar"218$}}
     \raise 2.0pt\hbox{$\mathchar"13C$}}}
\def\gta{\mathrel{\spose{\lower 3pt\hbox{$\mathchar"218$}}
     \raise 2.0pt\hbox{$\mathchar"13E$}}}

\newcommand{\de}[2]{\kern - #1 em \mathrm{d} #2}

\begin{document}

\title{Neutralino Dark Matter and the Curvaton}

\author{Martin Lemoine} \email{lemoine@iap.fr} \affiliation{Institut
d'Astrophysique de Paris, UMR 7095-CNRS, Universit\'e Pierre et Marie
Curie, 98bis boulevard Arago, 75014 Paris, France}

\author{J\'er\^ome Martin} \email{jmartin@iap.fr}
\affiliation{Institut d'Astrophysique de Paris, UMR 7095-CNRS,
Universit\'e Pierre et Marie Curie, 98bis boulevard Arago, 75014
Paris, France}

\date{\today}

\begin{abstract}
We build a realistic model of curvaton cosmology, in which the energy
content is described by radiation, WIMP dark matter and a curvaton
component. We calculate the curvature and isocurvature perturbations,
allowing for arbitrary initial density perturbations in all fluids,
following all species and their perturbations from the onset of dark
matter freeze-out onto well after curvaton decay. We provide detailed
numerical evaluations as well as analytical formulae which agree well
with the latter. We find that substantial isocurvature perturbations,
as measured relatively to the total curvature perturbation, can be
produced even if the curvaton energy density is well underdominant
when it decays; high precision measurements of cosmic microwave
background anisotropies may thus open a window on underdominant
decoupled species in the pre-nucleosynthesis early Universe. We also
find that in a large part of parameter space, curvaton decay produces
enough dark matter particles to restore WIMP annihilations, leading to
the partial erasure of any pre-existing dark matter - radiation
isocurvature perturbation.
\end{abstract}

\maketitle


\section{Introduction}

According to standard lore, the cosmological density perturbations at
the origin of large scale structure were seeded at some very early
time by the decay of the field that is also responsible for
inflation. In the past few years, there has been growing interest in a
variant of this cosmological scenario, in which another field, denoted
the ``curvaton'', communicates its own perturbations to matter and
radiation. In the simplest version, the curvaton plays a negligible
r\^ole in determining the dynamics of the Universe at the time of
inflation, but comes to dominate the energy density in the radiation
era, and as such, it becomes the seed of energy density perturbations
through its decay.  From a technical point of view, the curvaton
scenario transforms an initially isocurvature perturbation into an
adiabatic (or curvature) perturbation, as demonstrated in another
context in Ref.~\cite{M90} and more recently in Ref.~\cite{LW02} (see
also Ref.~\cite{BLO}).

\par 

In the ``simplest'' curvaton model, radiation is composed of a single
fluid deprived of an intrinsic density perturbation before curvaton
decay; the final perturbation is then purely adiabatic. However, as
already noted in~\cite{LW02}, radiation is generically a mixture of
different interacting fluids, so that if the curvaton decays after the
decoupling of one of the constituents, an isocurvature perturbation
will appear between this latter and the rest of radiation. Detailed
calculations have been provided in Refs.~\cite{LUW03,MT01} for
isocurvature perturbations generated between photons and baryons,
between photons and neutrinos as well as between photons and dark
matter and their impact on the CMB anisotropies. Non-gaussian
signatures are also to be expected~\cite{LM97,LW03}. The
implementation of a curvaton scenario in the framework of high energy
physics and supersymmetric theories has been proposed in several
studies, see for instance~\cite{KKM03}.

\par 

Since any pre-existing isocurvature perturbation between two
interacting fluids is erased on a short timescale if the fluids share
thermal equilibrium~\cite{W04}, an isocurvature perturbation can only
be generated when the curvaton decays after the decoupling of one of
these components. In this respect, the case of WIMP dark matter -
photon isocurvature perturbation appears particularly interesting.
Dark matter is indeed expected to decouple very early on from
radiation, hence this leaves significant room for the curvaton to
decay after decoupling and before big-bang nucleosynthesis (as
required by cosmology). This may be put in contrast with baryons and
neutrinos which decouple just before big-bang nucleosynthesis.

\par 

To our knowledge, there is no exhaustive survey of the curvature and
isocurvature perturbations generated in a ``realistic'' curvaton model
which takes into account the existence of radiation {\em and} dark
matter, the influence of the freeze-out of this latter fluid, the
influence of initial density perturbations in radiation and dark
matter and the different decay widths of curvaton into radiation and
dark matter. For instance, Ref.~\cite{LUW03} has computed the
isocurvature and curvature perturbations under specific assumptions on
the origin of dark matter, neglecting the effect of freeze-out and of
different branching ratios. The generation of a net isocurvature
fluctuation during dark matter freeze-out consecutive to the presence
of the curvaton has been computed in Ref.~\cite{LW03} but without
discussing the evolution through curvaton decay.  In~\cite{LV04}, an
initial density perturbation in the radiation fluid has been taken
into account but the presence of dark matter was
neglected. Ref.~\cite{GMW04}, building on the general formalism
developed in~\cite{MWU03}, has considered a three-fluid model of
curvaton decay, albeit assuming that dark matter fully originates in
the decay of the curvaton. Finally, Ref.~\cite{FRV05} has presented a
detailed numerical evaluation of the curvature and isocurvature
perturbations in a three-fluid model of curvaton decay and their
impact on CMB anisotropies, without however specifying the origin of
dark matter.

\par

Our present objective is to make progress along these lines and to
present a comprehensive model of curvaton - radiation and WIMP dark
matter cosmology, building on the above earlier studies. We will
present analytical and numerical calculations of the perturbations
produced, following in detail the freeze-out of equilibrium of dark
matter annihilations and curvaton decay, with non-vanishing initial
perturbations in the radiation - dark matter sector and in the
curvaton field. We focus on WIMP (neutralino) dark matter, since it
offers at present one of the best motivated candidates for dark
matter. It further allows one to set a specific framework for the
thermal history of the different fluids, without having to rely on the
details of the underlying particle physics models. The scenario, and
its cosmological consequences, would certainly be different for axion
like dark matter, whose scrutiny is postponed to future study.

\par

The paper is arranged as follows. In the following section, \ie
Sec.~\ref{Detailed three-fluid model}, we describe the three-fluid
model and establish the equations of motion both at the background
level, see Sec.~\ref{Equations for background quantities}, and at the
perturbed level, see Sec.~\ref{Equations for perturbed quantities}. In
Sec.~\ref{Curvature and isocurvature perturbations}, we solve these
equations by approximate analytical methods. In Sec.~\ref{Numerical
evaluation and discussion}, in order to assess the validity of our
approximations, we investigate the evolution of the background and of
the perturbations by means of numerical calculations. The physical
interpretation of our results is also discussed at length in this
section. Finally, in Sec.~\ref{Conclusion}, we present our
conclusions.  Some specific technical details are also presented in
three appendices at the end of the paper. In App.~\ref{app:perturb},
we briefly recall how the gauge-invariant quantities relevant to the
present work are defined and how their equations of motion can be
derived. In App.~\ref{app:junct}, we explain how to quantify the
influence of the curvaton perturbations during freeze-out [our
Eq.~(\ref{eq:LW})] using junction conditions. In
App.~\ref{app:adiabat}, we provide an extended discussion of the
theorem of Weinberg~\cite{W04} on the approach to adiabaticity for the
perturbations of interacting fluids in terms of gauge-invariant
quantities.

\section{Detailed three-fluid model}
\label{Detailed three-fluid model}

We are interested in computing the curvature and isocurvature
perturbations produced in a three-fluid model composed of radiation
(denoted by the subscript $\gamma$ in what follows), a curvaton
component (denoted $\sigma$) and dark matter (denoted by $\chi$). The
dark matter particles share thermal equilibrium with radiation until
their freeze-out of equilibrium after having turned
non-relativistic. We denote by $x_{\rm f}$ the ratio of mass $m_\chi$
to temperature at which freeze-out occurs: $x_{\rm f}\equiv
m_\chi/T_{\rm f}$. In the particular case of neutralino dark matter,
one generically finds $x_{\rm f}\simeq20$. Unless otherwise noted, we
use $x_{\rm f}=20$ as a fiducial value in what follows.

\par

We further assume that the curvaton is very weakly coupled to the
visible sector (radiation + dark matter) and that it decays into this
sector {\em after} dark matter freeze-out. Were it to decay earlier,
there would be no final isocurvature perturbation between $\chi$ and
$\gamma$, in virtue of the theorem of Weinberg on the approach to
adiabaticity of interacting fluids~\cite{W04}. This point is addressed
with greater scrutiny in Appendix~\ref{app:adiabat}, where it is
shown, in particular, to be in excellent agreement with numerical
computations. If the curvaton were to couple to the visible sector so
that it had at some point shared thermal equilibrium, any pre-existing
isocurvature perturbation between itself and radiation or dark matter
would likewise have been erased. Hence the above assumption is one of
non-triviality. It is furthermore consistent, in the sense that a late
decay is a generic consequence of a very weak coupling; moduli fields,
in particular, emerge as natural candidates in this curvaton scenario.

\par

We denote by $\Gamma_\sigma$ the total decay width of the curvaton,
$\Gamma_{\sigma\gamma}$ and $\Gamma_{\sigma\chi}$ its partial decay
widths into radiation and dark matter, respectively. Decay of the
curvaton occurs when $\Gamma_\sigma\approx H$, hence our previous
assumption corresponds to $\Gamma_\sigma \ll \Gamma_{\chi\chi\vert\rm
f}$, where the latter quantity $\Gamma_{\chi\chi\vert\rm f}\equiv
n_{\chi\vert \rm f}\langle\sigma_{\chi\chi}v\rangle$ denotes the
$\chi-\chi$ annihilation rate at freeze-out. The annihilation
cross-section is related to the time of freeze-out by the standard
formula~\cite{KT90}:
\begin{equation}
\langle \sigma_{\chi\chi} v\rangle\,\simeq\, \frac{g_*^{1/2}}{
    0.038g}{\rm e}^{x_{\rm f}}\frac{\sqrt{x_{\rm f}} }{m_\chi\mP}\, ,
\end{equation}
with $\mP=G^{-1/2}\simeq 1.2\times10^{19}\,$GeV, $g=2$ and $g_*\simeq
230$. In the following, we describe how the previous physical
situations can be modeled at the background and perturbative level.

\subsection{Equations for background quantities}
\label{Equations for background quantities}

In the non-relativistic era, and in the absence of dark matter
production from curvaton decay, the number density of dark matter
particles decreases due to expansion and annihilation, see
e.g.~\cite{KT90}:
\begin{equation}
\label{evoln}
\frac{{\rm d}n_\chi}{{\rm d}t}+3Hn_\chi=-\left \langle
\sigma_{\chi\chi}v\right\rangle \left(n_\chi^2-n_{\chi ,{\rm
eq}}^2\right)\, ,
\end{equation}
where $n_{\chi,{\rm eq}}$ is the density at thermal equilibrium
defined by
\begin{equation}
\label{defneq}
n_{\chi ,{\rm eq}}\equiv \frac{g}{2\pi ^2}\int _{m_\chi}^{\infty }
\frac{\left(E^2-m_\chi^2\right)^{1/2}E}{\exp(E/T)+1}{\rm d}E\simeq
\frac{g}{(2\pi )^{3/2}}m_\chi^3x^{-3/2}{\rm e}^{-x}\, ,
\end{equation} 
where $g$ is the number of spin states. The last equality assumes that
the dark matter particles have become non-relativistic, \ie  $x\equiv
m_\chi/T \gg 1$. Since $\rho_\chi \simeq m_\chi n_\chi$ in the
non-relativistic era, the equation of evolution for $\rho_\chi$ takes
a similar form. In order to take into account a source term arising
from curvaton decay, it suffices to add a term
$\Gamma_{\sigma\chi}\rho_{\sigma}$ in the r.h.s. of the above
equation. One thus arrives at
\begin{equation}
\label{evolrho}
\frac{{\rm d}\rho_\chi }{{\rm d}t}+3H\rho_\chi =-\frac{\left \langle
\sigma_{\chi\chi} v\right\rangle }{m_\chi}\left(\rho_\chi^2-\rho_{\chi
,{\rm eq}}^2\right)\,+\,\Gamma_{\sigma\chi}\rho_\sigma \, .
\end{equation}
Dark matter annihilations produce radiation, consequently the
evolution of the energy density in the radiation fluid is described by
a similar equation:
\begin{equation}
\frac{{\rm d}\rho_\gamma }{{\rm d}t}+4H\rho_\gamma =+\frac{\left
\langle \sigma_{\chi\chi} v\right\rangle
}{m_\chi}\left(\rho_\chi^2-\rho_{\chi ,{\rm
eq}}^2\right)\,+\,\Gamma_{\sigma\gamma}\rho_\sigma\, .
\end{equation}
Finally, the energy density of the curvaton component decreases
through expansion and decay:
\begin{equation}
\frac{{\rm d}\rho_\sigma }{{\rm d}t}+3H\rho_\sigma
=-\,\Gamma_{\sigma}\rho_\sigma\, ,
\end{equation}
where, since $\Gamma _{\sigma }=\Gamma _{\sigma \chi }+\Gamma _{\sigma
\gamma }$, one has conservation of the total energy density. Note that
we treat the curvaton as a non-relativistic fluid or as a coherent
scalar field oscillating in a quadratic potential.

\par

These equations, as well as the Friedmann equations, can be recast in
dimensionless forms in terms of the density parameters
$\Omega_\gamma$, $\Omega_\chi$, $\Omega_\sigma$, the Hubble factor
$H$, and the $e-$fold number $N$ (or equivalently the parameter
$x\equiv m_\chi/T$), which is defined in terms of the scale factor $a$
by ${\rm d}N\,\equiv\,{\rm d}\ln a\,=\,{\rm d}\ln x$:
\begin{eqnarray}
\label{background}
\frac{{\rm d}\Omega _{\chi }}{{\rm d}N} &=& \Omega _{\sigma
  }\frac{\Gamma_{\sigma\chi}}{H}+\Omega _{\chi }\Omega _{\gamma }
  -\frac{3\langle \sigma_{\chi\chi} v\rangle \mP^2}{8\pi
  }\frac{H}{m_\chi} \left(\Omega _{\chi }^2-\Omega _{\chi ,{\rm
  eq}}^2\right)\, ,\\ \frac{{\rm d}\Omega _{\gamma }}{{\rm d}N} &=&
  \Omega _{\sigma }\frac{\Gamma_{\sigma\gamma}}{H}+\Omega _{\gamma
  }\left(\Omega _{\gamma}-1\right)+\frac{3\langle \sigma_{\chi\chi}
  v\rangle \mP^2}{8\pi }\frac{H}{m_\chi} \left(\Omega _{\chi
  }^2-\Omega _{\chi ,{\rm eq}}^2\right)\, ,\\ \frac{{\rm d}\Omega
  _{\sigma }}{{\rm d}N} &=& \Omega _{\sigma }\left( \Omega
  _{\gamma}-\frac{\Gamma_{\sigma\gamma}+\Gamma_{\sigma\chi}}{H}\right)\,
  ,\\ \frac{{\rm d}H}{{\rm d}N} &=& -\frac{3H}{2}\left(1+\frac{\Omega
  _{\gamma }}{3}\right)\, .
\end{eqnarray}
We assume that radiation is always in a state of thermal equilibrium,
so that the parameter $x$ which enters the formula for $\rho_{\chi
,{\rm eq}}$ can be written in terms of the radiation energy density,
hence $H$ and $\Omega_{\gamma}$:
\begin{equation}
\label{defT}
x\,=\,\frac{m_\chi}{ T}\,=\,\left(\frac{\pi^2g_*}{ 30}\frac{8\pi}{
  3\mP^2}\right)^{1/4} m_\chi H^{-1/2}\Omega_\gamma^{-1/4}\ .
\end{equation}
For the sake of simplicity, we ignore any temperature dependence of
the function $g_*$. 

\subsection{Equations for perturbed quantities}
\label{Equations for perturbed quantities}

We now turn to the description of the equations of motion for the
perturbed quantities. In Appendix~\ref{app:perturb}, it is shown how
to find the corresponding expressions from the general perturbation of
a fully covariant expression for the annihilation and decay
terms. Using these results, one can write the evolution equations for
the gauge invariant dimensionless quantities
$\Delta_{(\alpha)}\,\equiv\,\Delta\rho_{(\alpha)}/\rho_{(\alpha)}$
with $\alpha=\gamma,\chi,\sigma$, and $\Phi$:
\begin{eqnarray}
\label{pertchi}
\frac{{\rm d}\Delta _{\chi }}{{\rm d}N} &=&
-\frac{\Gamma_{\sigma\chi}}{H}\frac{\Omega _{\sigma }}{\Omega _{\chi
}} \left(\Delta _{\chi }-\Delta _{\sigma }\right) -\frac32
\left(\Omega _{\sigma }\Delta _{\sigma }+\Omega _{\gamma }\Delta
_{\gamma } +\Omega _{\chi }\Delta _{\chi }\right)
-\Phi\left(3-\frac{\Gamma_{\sigma\chi}}{H}\frac{\Omega _{\sigma
}}{\Omega _{\chi }}\right) +\frac{3H\langle \sigma_{\chi\chi} v\rangle
  \mP ^2}{8\pi m_\chi}\frac{1}{\Omega _{\chi }} \nonumber \\ & &
\times \left[-\Phi \left(\Omega _{\chi }^2-\Omega _{\chi ,{\rm
      eq}}^2\right) -2\left(\Omega _{\chi }^2\Delta _{\chi }-\Omega
  _{\chi ,{\rm eq}}^2\Delta _{\rm eq}\right) +\left(\Omega _{\chi
  }^2-\Omega _{\chi ,{\rm eq}}^2\right)\Delta _{\chi }\right] \, , \\
\label{pertgam}\frac{{\rm d}\Delta _{\gamma }}{{\rm d}N} &=&
-\frac{\Gamma_{\sigma\gamma}}{H}\frac{\Omega _{\sigma }}{\Omega
  _{\gamma }} \left(\Delta _{\gamma }-\Delta _{\sigma }\right) -2
  \left(\Omega _{\sigma }\Delta _{\sigma }+\Omega _{\gamma }\Delta
  _{\gamma } +\Omega _{\chi }\Delta _{\chi }\right)
  -\Phi\left(4-\frac{\Gamma_{\sigma\gamma}}{H}\frac{\Omega _{\sigma
  }}{\Omega _{\gamma }}\right) +\frac{3H\langle \sigma_{\chi\chi}
  v\rangle \mP ^2}{8\pi m_\chi}\frac{1}{\Omega _{\gamma }} \nonumber
  \\ & & \times \left[\Phi \left(\Omega _{\chi }^2-\Omega _{\chi ,{\rm
  eq}}^2\right) +2\left(\Omega _{\chi }^2\Delta _{\chi }-\Omega _{\chi
  ,{\rm eq}}^2\Delta _{\rm eq}\right) -\left(\Omega _{\chi }^2-\Omega
  _{\chi ,{\rm eq}}^2\right)\Delta _{\gamma }\right] \, , \\
  \label{pertsig} \frac{{\rm d}\Delta _{\sigma }}{{\rm d}N} &=&
  -\frac32 \left(\Omega _{\sigma }\Delta _{\sigma }+\Omega _{\gamma
  }\Delta _{\gamma } +\Omega _{\chi }\Delta _{\chi }\right)
  -\Phi\left(3+\frac{\Gamma_{\sigma\gamma}+\Gamma_{\sigma\chi}}{H}\right)\,
  ,\\ \label{pertphi}\frac{{\rm d}\Phi}{{\rm d}N} &=& -\Phi-\frac12
  \left(\Omega _{\sigma }\Delta _{\sigma }+\Omega _{\gamma }\Delta
  _{\gamma } +\Omega _{\chi }\Delta _{\chi }\right) \, ,
\end{eqnarray}
where we have introduced the short-hand notation
\begin{equation}
\label{defdneq}
\Delta _{\rm eq}\equiv \frac14 \left(\frac32 +\frac{m}{T}\right)\Delta
_{\gamma }\, . 
\end{equation}
This formula follows from Eq.~(\ref{defneq}), writing the temperature
fluctuations in terms of $\Delta \rho _{\gamma }$ according to
Eq.~(\ref{defT}). Note also that, throughout this study, we assume
that $\langle\sigma _{\chi \chi }v\rangle$ is a constant, independent
of temperature, meaning that annihilations only occur through $s-$wave
interactions.

\par

It is also interesting to re-write these equations in the following form:
\begin{eqnarray}
\label{evolzetachi}
3\frac{{\rm d}}{{\rm d}N}\left(-\Phi +\frac13\Delta _{\chi }\right)
&=& -\frac{\Gamma_{\sigma\chi}}{H}\frac{\Omega _{\sigma }}{\Omega
_{\chi }} \left(\Delta _{\chi }-\Delta _{\sigma
}\right)+\Phi\frac{\Gamma _{\sigma \chi}}{H}\frac{\Omega _{\sigma
}}{\Omega _{\chi }} +\frac{3H\langle \sigma_{\chi\chi} v\rangle \mP
^2}{8\pi m_\chi}\frac{1}{\Omega _{\chi }} \nonumber \\ & & \times
\left[-\Phi \left(\Omega _{\chi }^2-\Omega _{\chi ,{\rm eq}}^2\right)
-2\left(\Omega _{\chi }^2\Delta _{\chi }-\Omega _{\chi ,{\rm
eq}}^2\Delta _{\rm eq}\right) +\left(\Omega _{\chi }^2-\Omega _{\chi
,{\rm eq}}^2\right)\Delta _{\chi }\right] \, , \\ 4 \frac{{\rm
d}}{{\rm d}N} \left(-\Phi +\frac14\Delta_{\gamma }\right) &=&
-\frac{\Gamma_{\sigma\gamma}}{H}\frac{\Omega _{\sigma }}{\Omega
_{\gamma }} \left(\Delta _{\gamma }-\Delta _{\sigma
}\right)+\Phi\frac{\Gamma _{\sigma \gamma }}{H}\frac{\Omega _{\sigma
}}{\Omega _{\gamma }} +\frac{3H\langle \sigma_{\chi\chi} v\rangle \mP
^2}{8\pi m_\chi}\frac{1}{\Omega _{\gamma }} \nonumber \\ & & \times
\left[\Phi \left(\Omega _{\chi }^2-\Omega _{\chi ,{\rm eq}}^2\right)
+2\left(\Omega _{\chi }^2\Delta _{\chi }-\Omega _{\chi ,{\rm
eq}}^2\Delta _{\rm eq}\right) -\left(\Omega _{\chi }^2-\Omega _{\chi
,{\rm eq}}^2\right)\Delta _{\gamma }\right] \, , \\ 3\frac{{\rm
d}}{{\rm d}N}\left(-\Phi +\frac13\Delta _{\sigma }\right) &=& -\Phi
\frac{\Gamma_{\sigma\gamma}+\Gamma_{\sigma\chi}}{H}\, .
\end{eqnarray}
When the r.h.s. of the previous expressions vanish and/or are
negligible, the equation of state of each component does not evolve in
time, and one finds that the quantities (where a dot means a
derivative with respect to cosmic time)
\begin{equation}
\zeta _{(\alpha)}\,\equiv\, -\Phi -H\frac{\Delta
  \rho_{(\alpha)}}{\dot \rho_{(\alpha)}}\, \simeq -\Phi
  +\frac{\Delta_{(\alpha)}}{3\left[1+\omega_{(\alpha)}\right]}\, ,
\label{eq:zeta}
\end{equation}
are conserved, a well-known result for isolated fluids. In this
expression, $\omega_{(\alpha)}=p_{(\alpha)}/\rho_{(\alpha)}$ is the
equation of state parameter for fluid $\alpha$.

\section{Curvature and isocurvature perturbations: analytical 
calculation}
\label{Curvature and isocurvature perturbations}

We now compute the isocurvature and curvature perturbations that arise
in this three-fluid model using analytical approximations for the
freeze-out of dark matter and for the decay of the curvaton. In the
next section, we compare the results of numerical evaluations of
the above equations to these analytical calculations.

\par

Since we assume that the curvaton never interacts with radiation nor
dark matter until it decays, $\zeta_\sigma$ is conserved (provided
that the equation of state of $\sigma$ does not change, which we
assume to be the case). Until dark-matter freeze-out, radiation and
dark matter share thermal equilibrium, hence $\zeta_\gamma =
\zeta_\chi$ as shown by Weinberg~\cite{W04}, see also
Appendix~\ref{app:adiabat}. Therefore, we can set the initial
conditions just prior to dark matter freeze-out, and write
$\zeta_{\sigma}^{\rm(i)}$, and
$\zeta_{\gamma}^{\rm(i)}=\zeta_{\chi}^{\rm(i)}$ the initial
gauge-invariant $\zeta_{(\alpha )}$ quantities in curvaton, radiation
and dark matter at that time as defined in Eq.~(\ref{eq:zeta}). We use
standard notations for the isocurvature modes, namely:
\begin{equation}
S_{\chi\gamma}\,\equiv\,3\left(\zeta_\chi-\zeta_\gamma\right),\quad
S_{\sigma\gamma}\,\equiv\,3\left(\zeta_\sigma-\zeta_\gamma\right)\ .
\end{equation}
Hence, initially $S_{\chi\gamma}^{\rm(i)}=0$ but
$S_{\sigma\gamma}^{\rm(i)}$ is generically non-zero.

\par

\begin{figure*}
  \centering
  \includegraphics[width=0.6\textwidth,clip=true]{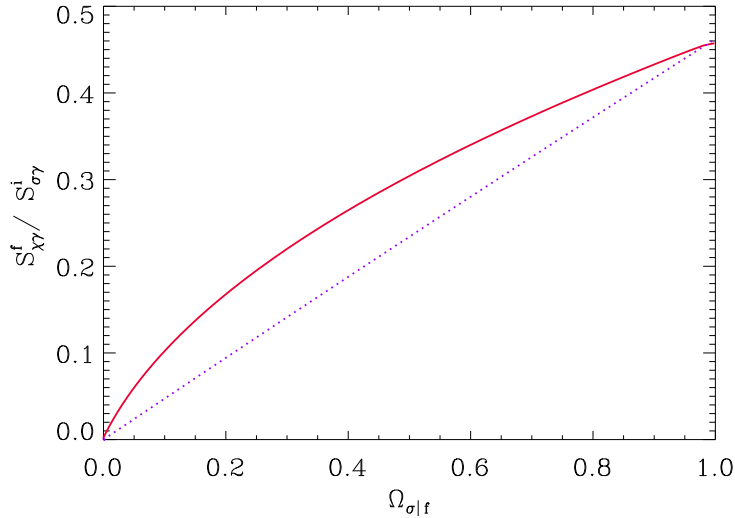}
  \caption[...]{Comparison of numerical (red solid line) and
  analytical (dotted blue line) calculations of the isocurvature
  perturbation $S_{\chi\gamma}$ generated by the curvaton during the
  freeze-out of $\chi-\chi$ annihilations, before curvaton decay.}
\label{fig:f2}
\end{figure*}

In the presence of a curvaton, a net isocurvature perturbation
$S_{\chi\gamma}$ may be generated at two different epochs: during dark
matter freeze-out, and at curvaton decay. Lyth and Wands~\cite{LW03}
have shown indeed that $S_{\chi\gamma}\neq0$ after freeze-out since
the very definition of freeze-out, $\Gamma_{\chi\chi}= H$ defines a
hypersurface which does not necessarily coincide with the hypersurface
of uniform radiation energy density. In other words, on this latter
hypersurface, dark matter and radiation do not present density
perturbations prior to freeze-out, yet freeze-out does not occur
at the same time at all space points as a result of the non-uniformity
of the Hubble factor on this hypersurface. In Appendix~\ref{app:junct}
we derive the resulting isocurvature perturbation using the properties
of junction conditions in general relativity. We obtain, in perfect
agreement with Lyth and Wands~\cite{LW03}:
\begin{equation}
\zeta _{\chi }^{>_{\rm f}}=\zeta _{\chi}^{<_{\rm
f}}+\frac{\left(\alpha_{\rm f}-3\right) \Omega _ \sigma^{>_{\rm f}}}
{2\left(\alpha _{\rm f}-2\right)+\Omega _ \sigma^{>_{\rm f}}}
\left(\zeta _{\sigma }^{<_{\rm f}}-\zeta _{\gamma }^{<_{\rm f}}\right)
=\zeta _{\chi }^{<_{\rm f}}+\frac13\frac{\left(\alpha_{\rm f}-3\right)
\Omega _ \sigma^{>_{\rm f}}} {2\left(\alpha _{\rm f}-2\right)+\Omega _
\sigma^{>_{\rm f}}} S_{\sigma \gamma }^{<_{\rm f}}\, ,
\label{eq:LW}
\end{equation}
where quantities denoted by a superscript $>_{\rm f}$ (resp. $<_{\rm
f}$) are evaluated just after (resp. just before) freeze-out. Notice
that we could also replace $\zeta _{\chi}^{<_{\rm f}}$ by $\zeta
_{\gamma}^{<_{\rm f}}$ because of the Weinberg theorem as discussed
before. The variable $\alpha$ is related to the scaling of the
annihilation rate with temperature; in practice, $\alpha_{\rm
f}=x_{\rm f}+3/2\sim20$. Freeze-out of annihilations does not
affect radiation nor the curvaton, hence one can safely assume:
$\zeta_\gamma^{>_{\rm f}}=\zeta_\gamma^{<_{\rm f}}$ and
$\zeta_\sigma^{>_{\rm f}}=\zeta_\sigma^{<_{\rm f}}$, as long as
curvaton decay has not yet occured. Therefore, a net isocurvature
perturbation $S_{\chi\gamma}$ is generated from the pre-existing
$S_{\sigma\gamma}$ curvaton -- radiation isocurvature perturbation:
\begin{equation}
S_{\chi\gamma}^{>_{\rm f}}\,=\, \frac{\left(\alpha_{\rm f}-3\right)
\Omega _ \sigma^{>_{\rm f}}}{2\left(\alpha _{\rm f}-2\right)+\Omega _
\sigma^{>_{\rm f}}}S_{\sigma\gamma}^{\rm (i)}\ .
\end{equation}
As expected, $S_{\chi\gamma}^{>_{\rm f}}$ vanishes in the limit
$\Omega_\sigma^{>_{\rm f }}\rightarrow 0$. In Fig.~\ref{fig:f2}, we
compare the above analytical approximation with detailed numerical
integration of the equations for the perturbed quantities that were
presented in the previous section. There is overall agreement although
the two curves tend to differ by a factor of a few at small values of
$\Omega_\sigma^{>_{\rm f}}$, the energy density parameter in the
curvaton at the time of freeze-out.

During curvaton decay, a net isocurvature perturbation
$S_{\chi\gamma}$ may also be produced if either dark matter or
radiation (but not both) inherits (part or all of) the curvaton
perturbation. This isocurvature perturbation can be calculated as
follows. In the time interval following dark matter freeze-out and
preceding curvaton decay, the various $\zeta_{(\alpha)}$ are
separately conserved since the various fluids do not interact with
each other. In order to compute the evolution of the curvature
perturbations through curvaton decay, one may use the total $\zeta$,
namely
\begin{equation} 
\zeta =\sum _{(\alpha )}\frac{\dot{\rho }_{(\alpha )}}{\dot{\rho
}}\zeta _{(\alpha )} =r\zeta _{\sigma }+s\zeta _{\gamma
}+\left(1-r-s\right)\zeta _{\chi }\, ,
\end{equation}
where 
\begin{equation}
\label{defrs}
r\equiv \frac{3\Omega _{\sigma }}{3\Omega _{\sigma }+3\Omega _{\chi }
+4\Omega _{\gamma }}\, \qquad s\equiv \frac{4\Omega _{\gamma
}}{3\Omega _{\sigma }+3\Omega _{\chi } +4\Omega _{\gamma }}\, .
\end{equation}
Although $\zeta$ evolves in time due to the evolution of $r$ and $s$,
it is conserved through decay if one assumes curvaton decay to be
instantaneous. Hence one can relate the quantities immediately after
decay, denoted by the superscript $>_{\rm d}$, to those immediately
before decay, denoted $<_{\rm d}$, through the relation:
\begin{equation}
r^{<_{\rm d}}\zeta_{\sigma}^{<_{\rm d}}+ s^{<_{\rm
d}}\zeta_{\gamma}^{<_{\rm d}}+ \left(1-r^{<_{\rm d}}-s^{<_{\rm
d}}\right)\zeta_{\chi }^{<_{\rm d}} \,=\, t^{>_{\rm
d}}\zeta_{\gamma}^{>_{\rm d}}+ (1-t^{>_{\rm d}})\zeta_{\chi}^{>_{\rm
d}}\ ,
\end{equation}
with
\begin{equation}
t\,=\,s\left(\Omega _{\sigma }=0\right)=\frac{4\Omega_\gamma}{
4\Omega_\gamma + 3\Omega_\chi}\ .
\end{equation}
We may further assume that the energy density contained in dark matter
immediately before and immediately after decay is negligible. Curvaton
decay must indeed proceed before big-bang nucleosynthesis, and dark
matter domination well after. Then we may set $t^{>_{\rm d}}\simeq 1$,
$(1-r^{<_{\rm d}}-s^{<_{\rm d}})\simeq 0$, so that:
\begin{equation}
\zeta_\gamma^{>_{\rm d}}\,=\,(1-r^{<_{\rm d}})\zeta_\gamma^{<_{\rm d}}
+ r^{<_{\rm d}}\zeta_\sigma^{<_{\rm d}} =\zeta_\gamma^{<_{\rm
d}}+\frac13r^{<_{\rm d}}S_{\sigma \gamma }^{<_{\rm d}} \,
.\label{eq:zetagammad}
\end{equation}
This result for $\zeta_\gamma$ had been obtained previously in
two-fluid models of the curvaton scenario. If one removes the
assumption $t^{>_{\rm d}}\simeq 1$ (which implicitly implies
$B_{\gamma }\equiv \Gamma_{\sigma\gamma}/\Gamma_{\sigma}\simeq 1$), it
is possible to show that, in the limit $B_{\gamma }\rightarrow 0$, one
finds $\zeta_\gamma^{>_{\rm d}}\rightarrow \zeta_\gamma^{<_{\rm d}}$
independently of $r^{<_{\rm d}}$, as it should. Moreover, the
coefficient $r^{<_{\rm d}}\propto \Omega _{\sigma }^{<_{\rm d}}$ and,
therefore, in the absence of a curvaton, the quantity $\zeta _{\gamma
}$ is unaffected as expected. Finally, if there is no pre-existing
isocurvature perturbation between the curvaton and radiation,
$\zeta_{\gamma }$ is also left unchanged as the perturbations share a
common origin.

\par

Let us also remark that, since $\zeta_\gamma^{<_{\rm d}}\simeq
\zeta_\gamma^{>_{\rm f}}\simeq\zeta_\gamma^{<_{\rm f}}\simeq
\zeta_{\gamma}^{\rm(i)}$ and similarly $\zeta_\sigma^{<_{\rm d}}\simeq
\zeta_{\sigma}^{\rm(i)}$, the above equation allows to express the
final perturbation in radiation in terms of the initial
perturbations. This will be done further below.

\par

In order to evaluate the final curvature perturbation in dark matter,
it is convenient to construct a compound fluid that comprises the dark
matter and a fraction $B_\chi\equiv\Gamma_{\sigma\chi}/\Gamma_\sigma$
of the curvaton fluid such that the source term coming from curvaton
decay vanishes in its equation of energy conservation~\cite{GMW04}:
\begin{equation}
\rho_{\rm comp}\,\equiv\,\rho_\chi + B_\chi \rho_\sigma\ .
\end{equation}
For this compound fluid, the curvature perturbation $\zeta_{\rm comp}$
is conserved, with:
\begin{equation}
\zeta_{\rm comp}\,=\,\frac{\Omega_\chi}{ \Omega_\chi +
  B_\chi\Omega_\sigma} \zeta_\chi + \frac{B_\chi\Omega_\sigma }{
  \Omega_\chi + B_\chi\Omega_\sigma} \zeta_\sigma\ .
\end{equation}
The conservation of $\zeta_{\rm comp}$ comes from the fact that the
source terms also cancel out at the perturbed level. Notice that this
is possible only because the curvaton and the dark matter have the
same equation of state; in particular, one could not construct a
simple compound fluid made of radiation and the non-relativistic
curvaton. Then, writing $\zeta_{\rm comp}^{>_{\rm d}}=\zeta_{\rm
comp}^{<_{\rm d}}$ and $\Omega_{\sigma}^{>_{\rm d}}=0$ leads to:
\begin{equation}
\zeta_\chi^{>_{\rm d}}\,=\,\frac{\Omega_\chi^{>_{\rm f}}}{
  \Omega_\chi^{>_{\rm f}} + B_\chi\Omega_\sigma^{>_{\rm f}}}
  \zeta_\chi^{>_{\rm f}} + \frac{B_\chi\Omega_\sigma^{>_{\rm f}}}{
  \Omega_\chi^{>_{\rm f}} + B_\chi\Omega_\sigma^{>_{\rm f}}}
  \zeta_\sigma^{>_{\rm f}} =\zeta_\chi^{<_{\rm d}}+\frac13
  \frac{B_\chi\Omega_\sigma^{>_{\rm f}}}{ \Omega_\chi^{>_{\rm f}} +
  B_\chi\Omega_\sigma^{>_{\rm f}}} S_{\sigma \chi }^{<_{\rm d}}\ .
\label{eq:zetachid}
\end{equation}
This equation has a structure very similar to
Eq.~(\ref{eq:zetagammad}). It indicates that dark matter perturbations
inherit a contribution from curvaton perturbations only if the
branching ratio $B_{\chi }$, the curvaton contribution before decay
and the pre-existing $S_{\sigma \chi }^{<_{\rm d}}$ do not vanish, as
is intuitively clear.

\par

Combining Eqs.~(\ref{eq:zetachid}) and (\ref{eq:zetagammad}), one may
thus write the isocurvature perturbation $S_{\chi\gamma}$ generated
through curvaton decay as a function of the pre-existing
$S_{\chi\gamma }$, $S_{\chi\sigma}$ and/or $S_{\sigma \gamma}$:
\begin{equation}
S_{\chi\gamma}^{>_{\rm d}}\,=\,S_{\chi\gamma}^{<_{\rm d}}+
\frac{B_\chi\Omega_\sigma ^{>_{\rm f}}}{ \Omega_\chi^{>_{\rm f}} +
B_\chi\Omega_\sigma^{>_{\rm f}}}S_{\sigma \chi}^{>_{\rm f}} -
r^{<_{\rm d}}S_{\sigma\gamma}^{<_{\rm d}}\ .
\end{equation}
In the r.h.s. of the above equation, terms can be evaluated at any time
in the interval from post-freeze-out to pre-decay, except $r$ which
must be calculated immediately before decay.

\par

To summarize, the final radiation curvature perturbation can be
expressed as
\begin{eqnarray}
\label{zetagf}
\zeta^{\rm (f)}_\gamma&\,=\,& \zeta_\gamma^{\rm (i)} + \frac{1}{
3}r^{<_{\rm d}}S_{\sigma\gamma}^{\rm (i)}\ .
\end{eqnarray} 
This final $\zeta^{\rm (f)}_\gamma$ is in fact nothing but $\zeta
_\gamma$ given by Eq.~(\ref{eq:zetagammad}) because the perturbations
in the radiation fluid (and in the curvaton) are not affected by the
freeze-out. On the other hand, using Eqs.~(\ref{eq:zetachid}) and
(\ref{eq:zetagammad}), the final dark matter curvature perturbations
can be written as the sum of two contributions, one from the
freeze-out and one from the curvaton decay, namely
\begin{eqnarray}
\zeta^{\rm (f)}_\chi &=& \zeta _{\chi }^{<_{\rm
f}}+\frac13\frac{\left(\alpha_{\rm f}-3\right) \Omega _\sigma^{>_{\rm
f}}}{2\left(\alpha _{\rm f}-2\right)+ \Omega _\sigma^{>_{\rm f}}}
S_{\sigma \gamma }^{<_{\rm f}}+ \frac13 \frac{B_\chi\Omega_\sigma^{>_{\rm
f}}}{\Omega_\chi^{>_{\rm f}} + B_\chi\Omega_\sigma^{>_{\rm f}}}
S_{\sigma \chi}^{<_{\rm d}} \nonumber\\ &=& \zeta _{\chi }^{<_{\rm
f}} +\frac13 \frac{\left(\alpha_{\rm f}-3\right) \Omega
_\sigma^{>_{\rm f}}}{2\left(\alpha _{\rm f}-2\right)+ \Omega
_\sigma^{>_{\rm f}}} \frac{\Omega_\chi^{>_{\rm f}}}
{\Omega_\chi^{>_{\rm f}} + B_\chi\Omega_\sigma^{>_{\rm f}}} S_{\sigma
\gamma }^{<_{\rm f}} +\frac13 \frac{B_\chi\Omega_\sigma^{>_{\rm f}}}{
\Omega_\chi^{>_{\rm f}} + B_\chi\Omega_\sigma^{>_{\rm f}}} S_{\sigma
\chi}^{<_{\rm f}}
\label{zetachif}
\end{eqnarray}
Therefore, combining Eqs.~(\ref{zetagf}) and (\ref{zetachif}), our
result for the final isocurvature perturbation in terms of the initial
isocurvature perturbation can be written as
\begin{eqnarray}
S^{\rm (f)}_{\chi\gamma}&\,=\,&\left[\frac{(\alpha_{\rm
  f}-3)\Omega_{\sigma}^{>_{\rm f}}}{ 2(\alpha_{\rm
  f}-2)+\Omega_{\sigma}^{>_{\rm f}}}\frac{\Omega_{\chi}^{>_{\rm f}}}{
  \Omega_{\chi}^{>_{\rm f}} + B_\chi\Omega_{\sigma}^{>_{\rm f}}} +
  \frac{B_\chi\Omega_{\sigma}^{>_{\rm f}}}{ \Omega_{\chi}^{>_{\rm f}}
  + B_\chi\Omega_{\sigma}^{>_{\rm f}}}-r^{<_{\rm
  d}}\right]S_{\sigma\gamma}^{\rm (i)}\ ,\label{eq:res}
\end{eqnarray}
where we have used the results: $\zeta _{\chi }^{<_{\rm f}}=\zeta
_{\gamma}^{({\rm i})}$ and $S_{\sigma\gamma}^{<_{\rm
f}}=S_{\sigma\chi}^{<_{\rm f}}=S_{\sigma\gamma}^{\rm (i)}$. Assuming
radiation domination all throughout freeze-out, the energy densities
just after freeze-out are given by [see Ref.~\cite{KT90}, especially
Eq.~(5.45)]\footnote{We compute $n_{\chi}^{>_{\rm f}}$ using the
variable $Y_\infty$ defined in this equation and assuming radiation
domination at freeze-out; this notably guarantees
$\Gamma_{\chi\chi}=H$ at freeze-out, which would not be obtained if
one used $n_{\rm eq}^{>_{\rm f}}$ instead.}:
\begin{equation}
\label{linkif}
\Omega_\sigma^{>_{\rm f}}\,=\,\Omega_\sigma^{\rm (i)}\frac{x_{\rm f}}{
  x_{\rm i}},\quad \Omega_\chi^{>_{\rm f}}\,\simeq\, 1.67 \times
  10^{-3}\,x_{\rm f}^{3/2}{\rm e}^{-x_{\rm f}}\ .
\end{equation}

\par

It proves interesting to discuss various limits of
Eq.~(\ref{eq:res}). Consider for instance the case $B_\chi
\Omega_\sigma^{>_{\rm f}}\gg \Omega_\chi^{>_{\rm f}}$. Physically,
this means that curvaton decay will produce more dark matter than
existed prior to decay, hence dark matter entirely inherits the
curvature perturbation of the curvaton: $\zeta_\chi^{\rm
(f)}\sim\zeta_\sigma^{\rm (i)}$ and consequently, $S_{\chi\gamma}^{\rm
(f)}\simeq (1-r^{<_{\rm d}})S_{\sigma\gamma}^{\rm (i)}$. In the limit
in which the curvaton density parameter is negligible at the time of
decay, $r^{<_{\rm d}}\ll 1$, only a negligible fraction of radiation
has been produced during curvaton decay, hence radiation conserves its
initial curvature perturbation: $\zeta_\gamma^{\rm (f)}\sim
\zeta_\gamma^{\rm (i)}$, and therefore the isocurvature perturbation
$S_{\chi\gamma}^{\rm (f)}$ is maximal.  We find it important to stress
that in this limit, {\em a large isocurvature perturbation may be
generated during curvaton decay, even if the curvaton contributes a
negligible amount to the total energy density at the time of its
decay}. This result comes about because the energy density of dark
matter in the very early Universe, prior to big-bang nucleosynthesis,
is itself very small, hence it can be strongly affected by the decay
of underdominant species.

\par

In the opposite limit, $r^{<_{\rm d}}\sim 1$, meaning
$\Omega_\sigma^{<_{\rm d}}\sim 1$, the decay of the curvaton
regenerates both radiation and dark matter, hence the final
isocurvature perturbation turns out negligible.

\par

It may also be that the branching ratio $B_\chi$ of curvaton to dark
matter is negligibly small, or even zero, for instance if the decay is
kinematically forbidden. In this case, a net isocurvature perturbation
may arise from two different sources: freeze-out and/or transfer of
$\sigma$ perturbations to radiation, as can be seen in
Eq.~(\ref{eq:res}). The former depends on the magnitude of the
curvaton density parameter at the time of freeze-out: it is negligible
if $\Omega_\sigma^{>_{\rm f}}\ll 1$, but tends to $\approx 50\,$\% as
$\Omega_\sigma^{>_{\rm f}}\rightarrow 1$. The latter increases with
$r^{<_{\rm d}}$, as it should. In particular, $S_{\chi\gamma}^{\rm
(f)}$ is maximal if $\Omega_\sigma^{>_{\rm f}}\ll 1$, so that
$\zeta_\chi^{\rm (f)}\sim \zeta_\chi^{\rm (i)}$, but $r^{<_{\rm
d}}\sim 1$, so that $\zeta_\gamma^{\rm (f)}\sim \zeta_\sigma^{\rm
(i)}$, \ie all of the curvaton perturbations is transfered to
radiation but none to dark matter.

\par

These limits will be encountered in the following Section, in which we
compare the above analytical results to numerical integration of the
equations derived in the previous section.

\section{Curvature and isocurvature perturbations: numerical 
evaluation and discussion}
\label{Numerical evaluation and discussion}

The above analytical calculations assume instantaneous dark matter
freeze-out and instantaneous curvaton decay. In order to assess the
validity of the previous results, we have integrated numerically the
equations of motion given in Section~\ref{Detailed three-fluid model},
starting before freeze-out at $x_{\rm i}=3$, until well after curvaton
decay. In general, the previous analytical formulae are found to be
accurate but they tend to overestimate the final isocurvature
perturbation in a substantial fraction of parameter space, by a factor
sometimes as large as an order of magnitude. The source of this error
lies in the neglect in the above analytical calculations of the
coupling between dark matter and radiation after freeze-out. Indeed,
when a significant amount of dark matter particles is produced by
curvaton decay, annihilations may become effective again, which leads
to a  erasure of the pre-existing $S_{\chi\gamma}$ isocurvature
perturbation. This will be detailed and quantified in what follows.
The numerical integration also allows to follow the evolution of the
curvature and isocurvature perturbations in the different components
in parallel with their abundance, which as we will show, offers a
sound and intuitive understanding of their behavior.

\subsection{Background and perturbations evolution}

The parameters that control this evolution are obviously:
$\Omega_\sigma^{\rm (i)}$, which controls the relative magnitude of
the curvaton energy density at decay and at freeze-out,
$\Gamma_{\sigma}$ (as measured relative to the fixed quantity
$\Gamma_{\chi\chi\vert\rm f}$) which controls the time of decay, the
branching ratio $B_{\chi}=\Gamma_{\sigma\chi}/\Gamma_{\sigma}$, which
controls the amount of dark matter produced relative to radiation
during curvaton decay, and
$\zeta_\gamma^{\rm(i)}/\zeta_\sigma^{\rm(i)}$, the ratio of initial
density perturbations in radiation (or dark matter) and the
curvaton. However, the value of
$\zeta_\gamma^{\rm(i)}/\zeta_\sigma^{\rm(i)}$ only influences the
final total curvature perturbation $\zeta_{\gamma}^{\rm(f)}$ [or
$\zeta_{\chi}^{\rm(f)}$] as expressed relatively to
$\zeta_\sigma^{\rm(i)}$, see Eq.~(\ref{eq:res}). The isocurvature
perturbation $S_{\chi\gamma}^{\rm(f)}/S_{\sigma\gamma}^{\rm(i)}$, \ie
measured relatively to the initial isocurvature perturbation, does not
depend on this parameter, see Eq.~(\ref{eq:res}). We may therefore
keep $\zeta_\gamma^{\rm(i)}/\zeta_\sigma^{\rm(i)}$ to a fixed value
(here, $0.31$) 
without affecting the generality of our results.

 In principle, one could determine these various parameters if the
couplings of the curvaton to the differents sectors of the theory as
well as the early dynamics of the curvaton {\em and} the inflaton
and their perturbed components during the inflationary era were
specified. This however lies well beyond the goals of the present
study. Furthermore, our objective is to present a complete survey of the
various alternative cosmological effects, hence we have chosen to
select reasonable  values for these parameters in order to bring to light
the most relevant consequences.

The time of freeze-out, encoded in $x_{\rm f}$, also affects the
discussion since it directly determines the dark matter energy
density. We will keep this value fixed to the generic 
$x_{\rm f}=20$ in most of our discussion, but discuss its influence,
and show one particular case of the evolution in the case $x_{\rm
f}=30$. We have also decided to keep the time of radiation-matter
equality free although it should of course be fixed to $T_{\rm
equ}\simeq 5.5 \Omega _0h^2\,$eV ($x_{\rm equ}\sim
10^{11}$)~\cite{KT90}. It is always possible to achieve the right dark
matter abundance by tuning the above parameters, but this would be
obtained at the price of unnecessary complexity. However, we set the
time of curvaton decay to take place before big-bang nucleosynthesis,
which starts at $x_{_{\rm BBN}}\simeq 10^5 (m_\chi/100\,{\rm GeV})$.

\par

In order to survey the various effects in this extended parameter
space, we plot in a first series of figures typical cases of the
evolution of the background and perturbed variables as a function of
$e-$foldings of the scale factor (actually $x\equiv m_{\chi}/T$) for
various values of $\Omega_\sigma^{\rm (i)}$ and $B_\chi$. 

\begin{figure*}
  \centering
  \includegraphics[width=0.6\textwidth,clip=true]{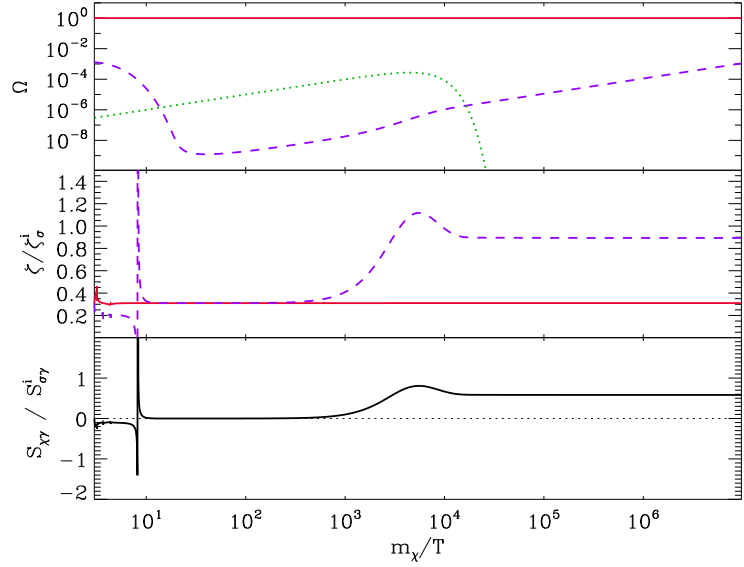}
  \caption[...]{Evolution of background quantities (upper panel)
$\Omega_\chi$ (dashed line), $\Omega_\gamma$ (solid line),
$\Omega_\sigma$ (dotted line), curvature perturbations (middle panel)
$\zeta_\gamma/\zeta_\sigma^{\rm(i)}$ (solid line) and
$\zeta_\chi/\zeta_\sigma^{\rm(i)}$ (dashed line), and the isocurvature
perturbation $S_{\chi\gamma}^{\rm(f)}/S_{\sigma\gamma}^{\rm(i)}$
(lower panel). Parameters are: $B_\chi=10^{-3}$,
$\Gamma_{\sigma}=10^{-6}\Gamma_{\chi\chi\vert\rm f}$, and $\Omega_\sigma^{\rm
(i)}=3\times10^{-7}$ corresponding to $\Omega_\sigma^{<_{\rm d}}=
2.7\times10^{-4}$.  }
\label{fig:f3a}
\end{figure*}

  In Figure~\ref{fig:f3a}, the evolution of $\Omega_\chi$
(upper panel, dashed line), $\Omega_\gamma$ (upper panel, solid line),
$\Omega_\sigma$ (upper panel, dotted line),
$\zeta_\gamma/\zeta_\sigma^{\rm(i)}$ (middle panel, solid line),
$\zeta_\chi/\zeta_\sigma^{\rm(i)}$ (middle panel, dashed line), and
$S_{\chi\gamma}/S_{\sigma\gamma}^{\rm(i)}$ (lower panel), is displayed
as a function of $m_\chi/T$ for an initial value $\Omega _{\sigma
}^{\rm (i)}=3\times 10^{-7}$, which corresponds to $\Omega _{\sigma
}^{>_{\rm f}}= 2.0\times 10^{-6}$ [see Eq.~(\ref{linkif})] and to a
``small'' value of $\Omega_\sigma^{<_{\rm d}}\,=\,2.7\times10^{-4}$,
that is to say a small value of $r^{<_{\rm d}}$. The branching ratio
is taken to be $B_\chi =10^{-3}$; the curvaton decay width is set to
$\Gamma_\chi=10^{-6}\Gamma_{\chi\chi\vert\rm f}$, which corresponds to
a time of decay $x_{\rm d}=4.7\times 10^3$ (defined as the time at
which $H=\Gamma_\chi$), or, in terms of temperature, $T^{>_{\rm
d}}\simeq 21\,$MeV (assuming $m_\chi=100\,$GeV).

 As shown in the upper panel of this figure, the Universe remains
radiation dominated at all times before radiation-dark matter
equality. The curvaton energy density increases relative to that of
radiation in proportion to $a(t)\propto x$, which is characteristic of
non-relativistic matter, then vanishes exponentially as the curvaton
decays. The dark matter energy density parameter first decreases
exponentially due to annihilations in the non-relativistic regime,
then stagnates when freeze-out occurs at $x_{\rm f}=20$ and finally
increases linearly with $a$, just as the curvaton, in the
post-freeze-out era. The curvaton decay produces a significant number
of dark matter particles at $x\simeq0.5-1.0\times 10^4$, which
translates in a faster increase of $\Omega_\chi$ in this region.

 The middle panel in this figure illustrates the concomittant
evolution of the curvature perturbations $\zeta_\alpha$ in these
components. The spikes that appear for $\zeta_\gamma/\zeta_\sigma^{\rm
i}$ and $\zeta_\chi/\zeta_\sigma^{\rm i}$ in the pre-freeze-out era
come about because the curvature perturbations involve the inverses of
the time derivatives of $\rho_\gamma$ and $\rho_\chi$, see
Eq.~(\ref{eq:zeta}), which may become singular locally when
production/annihilation terms are present in the equations of motion ;
this, nevertheless, does not signify that perturbation becomes
inapplicable but rather that the spatial hypersurfaces are ill-defined
at that point~\cite{MWU03}.

 Since the curvaton and dark matter are so much underabundant with
respect to radiation, this latter fluid is isolated in a good
approximation, hence its curvature perturbation $\zeta_\gamma$ remains
conserved, as seen in the middle panel.  In contrast, the dark matter
perturbation deviates from its initial value at $x\sim 10^3-10^4$ as
it acquires the curvaton perturbations through the decay of this
latter. The radiation-dark matter isocurvature perturbation then
becomes non-vanishing.

 The physical situation depicted in this figure is particularly
interesting because it confirms that a {\em significant isocurvature
perturbation
$S_{\chi\gamma}^{\rm(f)}/S_{\sigma\gamma}^{\rm(i)}\simeq90\,$\% is
generated even though the curvaton field never contributes
significantly to the total energy density content}. This stems from
the very small energy density of the dark matter component at the time
of curvaton decay relative to the amount of energy transfer from
curvaton to dark matter at that time ($B_\chi\Omega_\sigma^{<_{\rm
d}}$). Equivalently, in terms of these quantities evaluated at the
time of freeze-out: $B_\chi\Omega_\sigma^{>_{\rm f}}\sim 2\times
10^{-9}> \Omega_\chi^{>_{\rm f}}\sim 3\times 10^{-10}$, see
Eq.~(\ref{linkif}).  Accordingly, one can see $\zeta_{\chi}$ deviate
from its initial value at the time of curvaton decay in
Fig.~\ref{fig:f3a} and the value of $S_{\chi\gamma}$ evolves
simultaneously.

\begin{figure*}
  \centering
  \includegraphics[width=0.6\textwidth,clip=true]{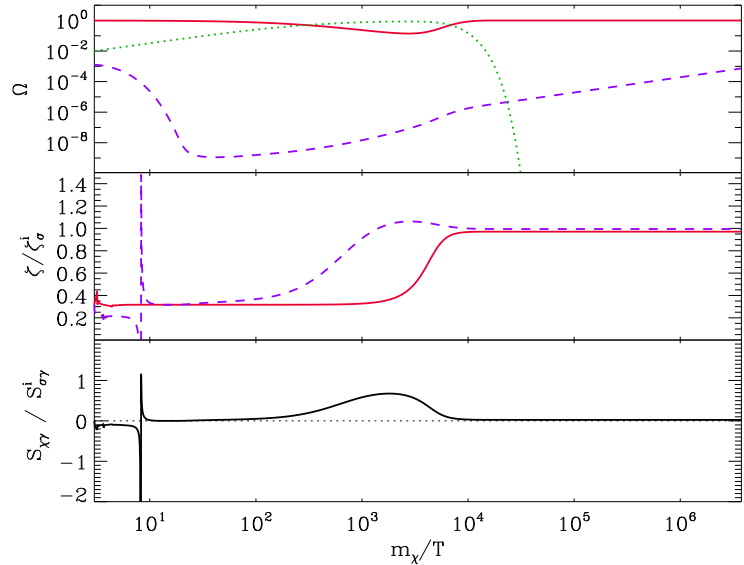}
  \caption[...]{Same as Fig.~\ref{fig:f3a} for the following
  parameters: $B_\chi=10^{-6}$,
  $\Gamma_{\sigma}=10^{-6}\Gamma_{\chi\chi\vert\rm f}$ and $\Omega_\sigma^{\rm
  (i)}=10^{-2}$ corresponding to $\Omega_\sigma^{<_{\rm d}}=0.65$.  }
\label{fig:f3b}
\end{figure*}

Figure~\ref{fig:f3b} shows a similar case, but with a higher initial
curvaton energy density, $\Omega _{\sigma }^{\rm (i)}=10^{-2}$, and a
smaller branching ratio to dark matter ($B_{\chi }=10^{-6}$). This
implies $\Omega_\sigma^{>_{\rm f}}= 6.7 \times 10^{-2}$ and
$\Omega_\sigma^{<_{\rm d}}=0.65$. 

 The evolution is similar to that seen in the previous figure,
although in the present case, the curvaton comes to dominate the
energy density before it decays, which implies that $\Omega_\gamma$
decreases somewhat during this period, before being regenerated during
curvaton decay. The middle panel shows that the curvature perturbation
of dark matter tends toward that of the curvaton well before decay at
$x\sim10^2-10^3$. Indeed, the time at which $\zeta_\chi$ departs from
its initial value is set by matching the amount of dark matter
produced by curvaton decay with that initially present. Furthermore
the energy density in the curvaton is larger here than in the previous
figure, hence this equality occurs earlier. Since the limit
$B_\chi\Omega_\sigma^{>_{\rm f}}> \Omega_\chi^{>_{\rm f}}$ remains
valid, a net isocurvature sets up, but gets erased as radiation is in
turn regenerated by curvaton decay ($r^{<_{\rm d}}\sim 1$ in this
case).

\begin{figure*}
  \centering
  \includegraphics[width=0.6\textwidth,clip=true]{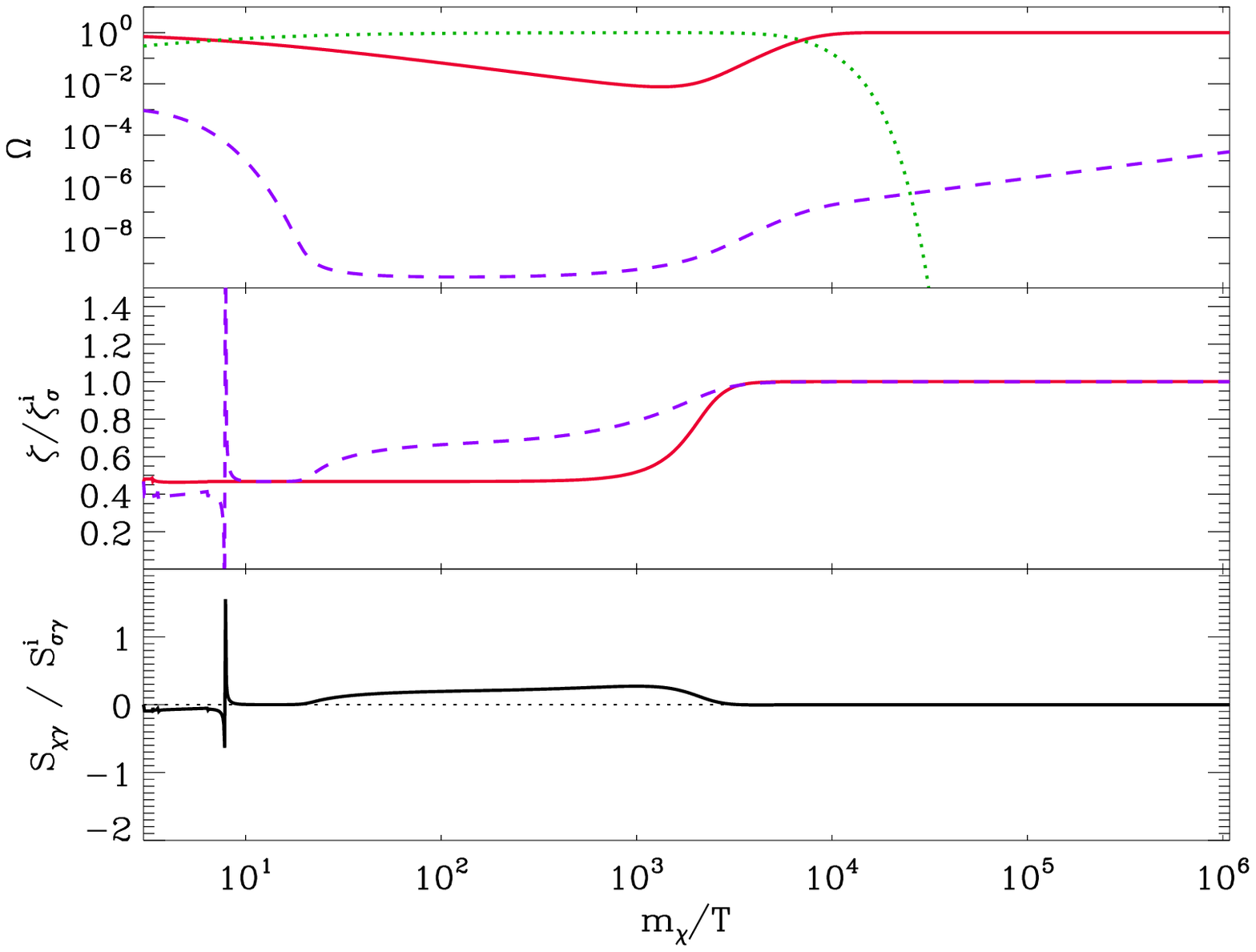}
  \caption[...]{Same as Fig.~\ref{fig:f3a} for the following
  parameters: $B_\chi=10^{-7}$,
  $\Gamma_{\sigma}=10^{-6}\Gamma_{\chi\chi\vert\rm f}$ and $\Omega_\sigma^{\rm
  (i)}=3\times10^{-1}$ corresponding to $\Omega_\sigma^{<_{\rm d}}=0.66$.  }
\label{fig:f3c}
\end{figure*}

 In Fig.~\ref{fig:f3c}, $\Omega _{\sigma }^{\rm (i)}=3\times 10^{-1}$
so that the curvaton dominates the energy density from $x\simeq 10$ to
$x\simeq 6\times10^4$, at which time it decays. Correspondingly, the
dark matter energy density parameter remains constant in the
post-freeze era, until it becomes sourced by curvaton decay at
$x\simeq10^3$; the radiation energy density parameter decreases in
proportion to $1/a(t)\propto 1/x$ until curvaton decay. As
$\Omega_\sigma^{\rm (i)}$ increases, its influence during freeze-out
becomes noticeable, see Fig.~\ref{fig:f3c} where $\Omega _{\sigma
}^{\rm (i)}=3\times 10^{-1}$ and $B_{\chi }=10^{-7}$. The isocurvature
perturbation $S_{\chi\gamma}$ is then generated at freeze-out due to
the effect discussed by Lyth \& Wands~\cite{LW03}.  Closer to curvaton
decay, the dark matter perturbation tends further toward that of the
curvaton. However, as in the previous example,
$B_\chi\Omega_\sigma^{>_{\rm f}}> \Omega_\chi^{>_{\rm f}}$, $r^{<_{\rm
d}}=\Omega_\sigma^{<_{\rm d}}\sim 1$ so that the isocurvature
perturbation gets erased as the radiation perturbation catches up with
that of dark matter.

\begin{figure*}
  \centering
  \includegraphics[width=0.6\textwidth,clip=true]{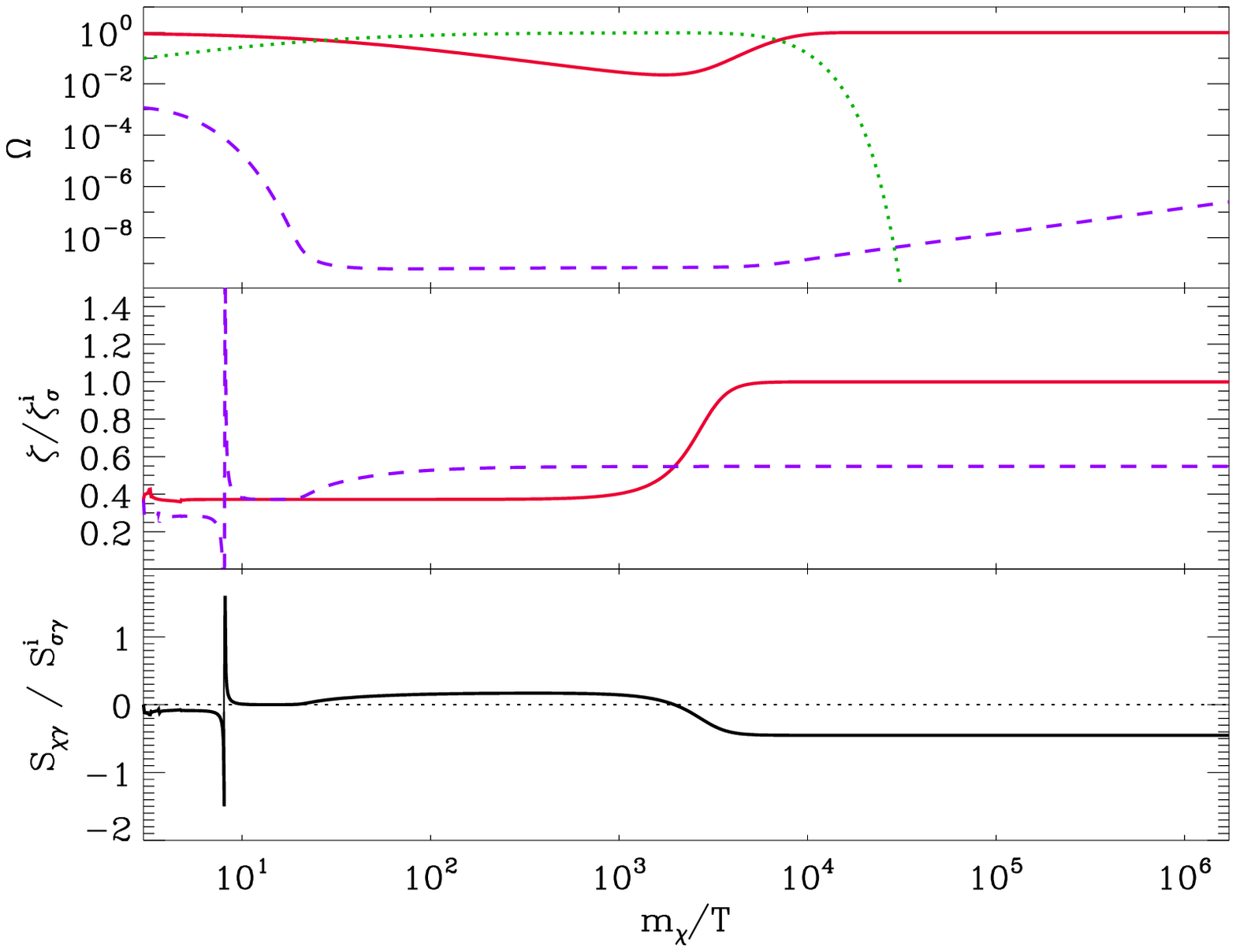}
  \caption[...]{Same as Fig.~\ref{fig:f3a} for the following
  parameters: $B_\chi=0$, $\Gamma_{\sigma}=10^{-6}\Gamma_{\chi\chi\vert\rm f}$
  and $\Omega_\sigma^{\rm (i)}=10^{-1}$ corresponding to
  $\Omega_\sigma^{<_{\rm d}}=0.66$.  }
\label{fig:f3d}
\end{figure*}

Finally, Fig.~\ref{fig:f3d} provides an example of the case in which
the decay of curvaton to dark matter is forbidden, \ie
$B_\chi=0$. Since $\Omega_\sigma^{<_{\rm d}}\sim 1$, all of radiation
is produced during the decay. However, dark matter nevertheless
inherits part of the curvaton perturbation through the influence of
this latter on the freeze-out, as is apparent at times $x\simeq
30-10^3$ in Fig.~\ref{fig:f3d}. Therefore the final $S_{\chi\gamma}$
isocurvature perturbation is not maximal, but of the order of 70\%, as
measured relatively to the initial $S_{\sigma\gamma}$.

\begin{figure*}
  \centering \includegraphics[width=0.6\textwidth,clip=true]{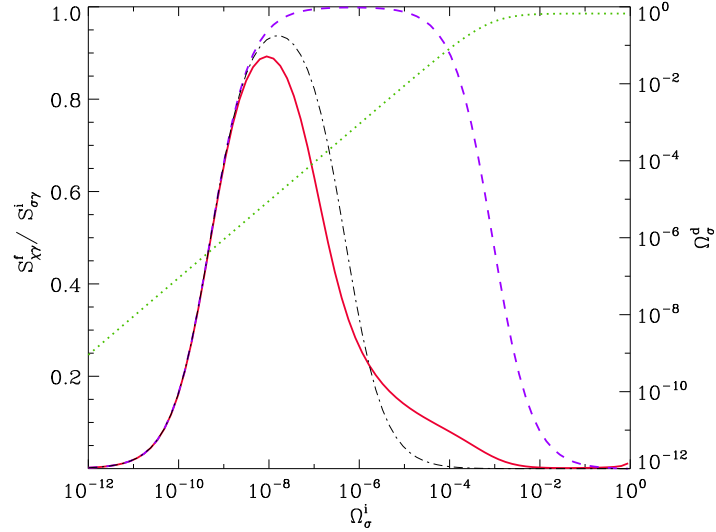}
  \caption[...]{Final isocurvature perturbation $S_{\chi\gamma}$
  expressed relatively to the initial $S_{\sigma\gamma}$ isocurvature
  perturbation (left $y-$axis), as a function of $\Omega_\sigma^{\rm
  (i)}$: numerical calculation (solid line), the analytical
  approximation given by Eq.~(\ref{eq:res}) (dashed line) and a
  refined approximation taking into account the late time annihilation
  due to the curvaton decay (dotted-dashed line). The dotted line
  indicates the corresponding value of $\Omega_\sigma^{<_{\rm d}}$
  (right $y-$axis) for convenience. The parameters are chosen to be:
  $\Gamma_\sigma/\Gamma_{\chi\chi\vert\rm f}=10^{-6}$ and $B_\chi=10^{-1}$.}
\label{fig:f4a}
\end{figure*}

\subsection{Final isocurvature perturbation}

In a second series of figures, we explore the dependence of
$S_{\chi\gamma}^{\rm(f)}/S_{\sigma\gamma}^{\rm(i)}$ on the initial
abundance of the curvaton (or, equivalently, its abundance at the time
of decay, $\Omega_\sigma^{<_{\rm d}}$) for different values of the
branching ratio $B_\chi$ of curvaton to dark matter, and compare the
numerical results to the analytical formulae.

\par

In Fig.~\ref{fig:f4a}, the branching ratio is chosen as
$B_\chi=0.1$. Let us describe this plot in more details. For very small
values of $\Omega_\sigma^{({\rm i})}$, there is no isocurvature
perturbation (as expected). This is also the case for large values of
$\Omega_\sigma^{({\rm i})}$, \ie when dark matter and radiation mostly
originate in curvaton decay and, hence, have a common
origin. Moreover, from our previous discussion of the analytical
calculations, one expects that $S_{\chi\gamma}^{\rm(f)}$ will be
maximal as long as $B_\chi\Omega_\sigma^{>_{\rm f}} >
\Omega_\chi^{>_{\rm f}}$ but $(1-B_\chi)\Omega_\sigma^{<_{\rm d}}\ll
\Omega_\gamma^{<_{\rm d}}\sim1$; this region corresponds to the peak
plateau of the dashed line in Fig.~\ref{fig:f4a}. Indeed, combining
Eqs.~(\ref{linkif}), the condition $B_\chi\Omega_\sigma^{>_{\rm f}}
\gg \Omega_\chi^{>_{\rm f}}$ reads:
\begin{equation}
B_\chi \Omega_\sigma^{\rm (i)}\gg 5 \times10^{-3}x_{\rm
  f}^{1/2}{\rm e}^{-x_{\rm f}}\ ,\label{eq:cond1}
\end{equation}
with $x_{\rm f}^{1/2}\exp(-x_{\rm f})\simeq 0.92\times10^{-8}$ for
$x_{\rm f}=20$ while the condition $(1-B_\chi)\Omega_\sigma^{<_{\rm
d}}\ll 1$ can be written as:
\begin{equation}
(1-B_\chi)\Omega_\sigma^{\rm
  (i)}\left(\frac{\Gamma_\sigma}{\Gamma_{\chi\chi\vert {\rm
  f}}}\right)^{-1/2}\ll 3 x_{\rm f}^{-1}\ ,\label{eq:cond2}
\end{equation}
since one assumes radiation domination all throughout and $x_{\rm
i}=3$. Recall that the ratio $\Gamma_\sigma/\Gamma_{\chi\chi\vert {\rm
f}}$ represents the ratio of the Hubble factors at curvaton decay and
at freeze-out.

As an example, consider the case depicted in Fig.~\ref{fig:f4a}:
$x_{\rm f}=20$, $\Gamma_\sigma/\Gamma_{\chi\chi\vert\rm f}=10^{-6}$ and
$B_\chi=10^{-1}$. Condition (\ref{eq:cond1}) shows that significant
isocurvature perturbation can be produced when $\Omega_\sigma^{\rm
(i)}\gtrsim 4.6\times10^{-10}$, which agrees well with the results
shown in Fig.~\ref{fig:f4a}. Condition (\ref{eq:cond2}) further limits
this region to $\Omega_\sigma^{\rm (i)}\lesssim 1.7\times10^{-4}$,
which, again, agrees with the limit of the dashed line (analytical
calculation).

\par

Despite the above considerations, one notices in Fig.~\ref{fig:f4a}
that, when $S_{\chi\gamma}^{\rm(f)}/S_{\sigma\gamma}^{\rm(i)}$ is
significant and $\Omega_\sigma^{\rm (i)}$ not too small, there is a
mismatch between the analytical and numerical results by a factor of a
few to an order of magnitude over most of the above parameter
space. The reason of this discrepancy has been alluded to earlier; it
lies in the neglect of the annihilation term in the post-freeze-out
era. If a sufficient number of dark matter particles are produced
during curvaton decay, this term may become significant, even at
values $x\gg x_{\rm f}$, and regenerates some form of coupling between
dark matter and radiation, leading to the partial or nearly complete
erasure of the isocurvature perturbation. One can actually observe a
typical example of the evolution of $S_{\chi\gamma}$ under the
influence of regenerated dark matter annihilations in
Fig.~\ref{fig:f3a} at times $x\simeq 10^{3.6}\rightarrow 10^4$: the
isocurvature perturbation decreases slightly to achieve its final
value during this time interval; the impact of annihilations on
$\Omega_\chi$ is not apparent in this figure but becomes clear if one
plots $\Omega_\chi/a\propto a^3n_\chi$. Note that dark matter
annihilations at around the time of big-bang nucleosynthesis may have
interesting phenomenological consequences and distinct astrophysical
signatures, see notably Ref.~\cite{J04}.

\par

One can assess the range of parameters in which the annihilation term
becomes significant, hence the analytical calculation inaccurate, as
follows. If curvaton decay is instantaneous and all of dark matter
originates in this decay (so as to produce a substantial
$S_{\chi\gamma}$), one has: $ \Omega_\chi^{>_{\rm
d}}\,\simeq\,B_\chi\Omega_\sigma^{<_{\rm d}} $. After curvaton decay,
one must compare the influence of the terms scaling with
$\langle\sigma_{\chi\chi}v\rangle$ in Eq.~(\ref{pertchi}) with the
terms due to expansion. Since $\Delta_\chi$ and $\Phi$ are of the same
order of magnitude, it suffices to compare the prefactors;
furthermore, $\Omega_{\chi ,{\rm eq}}\ll\Omega_\chi$ when $x\gg x_{\rm
f}$, so that, finally, the analytical calculation should be valid
provided:
\begin{equation}
\label{eq:cc1}
\frac{3H\langle\sigma_{\chi\chi}v\rangle \mP^2}{
8\pi   m_\chi}\Omega_\chi  \ll 1\ .
\end{equation}
Of course, it is understood that all terms in this equation should be
evaluated immediately after curvaton decay. Since $\Omega_\chi =
n_\chi m_\chi / \rho$, using the Friedmann equation one can rewrite
the above constraint in the more intuitive way:
\begin{equation}
\Upsilon\,\equiv\,\frac{\Gamma_{\chi\chi}^{>_{\rm d}}}{H^{>_{\rm
d}}}\,\ll\, 1\ ,
\label{eq:cond-up}
\end{equation}
which indeed expresses the fact that dark matter annihilations do not
occur since the annihilation rate remains smaller than the Hubble
rate. Note also that, in this sudden decay approximation,
annihilations cannot occur in the interval separating freeze-out from
curvaton decay.

\par

Alternatively, one can discuss the case in which curvaton decay is not
instantaneous, assuming here as well that $ \Omega_\chi^{>_{\rm
d}}\,\simeq\,B_\chi\Omega_\sigma^{<_{\rm d}} $. One must now compare
the influence of the terms scaling with
$\langle\sigma_{\chi\chi}v\rangle$ in Eq.~(\ref{pertchi}) with the
term due to curvaton decay, which leads to the following condition:
\begin{equation}
\label{importantcc}
\frac{3H\langle\sigma_{\chi\chi}v\rangle \mP^2}{
8\pi   m_\chi}\Omega_\chi ^2 \ll B_\chi \frac{\Gamma_\sigma}{
  H}\Omega_\sigma\ ,
\end{equation}
at all times before curvaton decay. By considering the time evolution
of the various terms, one can convince oneself that the l.h.s.
increases faster with $x$ than the r.h.s. Let us discuss this last
point in greater detail. The r.h.s. scales as $\propto x^3$ since
$\Omega _{\sigma }\propto x$ and $H\propto x^{-2}$ in a radiation
dominated phase. After freeze-out, but before curvaton decay, one
might think that, in Eq.~(\ref{background}), only the term $\Omega
_{\chi }\Omega _{\gamma }$ matters. In this case, $\Omega _{\chi
}\propto x$ and the l.h.s. is constant, \ie does not increase faster
than the r.h.s. However, the influence of curvaton decay is felt
before the time $H=\Gamma _{\sigma }$ so that the term proportional to
$\Omega _{\sigma }$ in Eq.~(\ref{background}) cannot always be
ignored.  When this term dominates, $\Omega _{\chi }\propto x^4$ hence
the l.h.s. indeed increases faster than the r.h.s. This scaling will
last as long as the annihilation term remains negligible. When this
latter becomes important, the evolution becomes such that the
annihilation term and the term proportional to $\Omega _{\sigma }$
balance each other, thus scaling in the same way, which leads to
$\Omega _{\chi }\propto x^{5/2}$ after the period where $\Omega _{\chi
}\propto x^4$ and until $H=\Gamma _{\sigma }$\footnote{In fact,
neglecting the term proportional to $\Omega _{\chi ,{\rm eq}}$,
Eq.~(\ref{background}) can be solved exactly. The solution reads
\begin{equation}
\Omega _{\chi }=\Omega _{\chi }^{>_{\rm f}} \frac{-\alpha
u^{2}I_1(v)+\alpha u^{5/2}\Upsilon ^{1/2}I_0(v) -\beta
u^{2}K_1(v)-\beta u^{5/2}\Upsilon ^{1/2}K_0(v)}{\left[\alpha
I_1(u)+\beta K_1(u)\right]}\, ,
\end{equation}
where $I$ and $K$ are modified Bessel functions and $u\equiv x/x_{\rm
f}$, $v\equiv 2\Upsilon ^{1/2}u^{1/2}$ [$\Upsilon$ being defined in
Eq.~(\ref{eq:cond-up})]. The quantities $\alpha $ and $\beta $ are two
constants that should be chosen such that $\Omega _{\chi }=\Omega
_{\chi }^{>_{\rm f}}$ when $u=1$ and such that the derivative of
$\Omega _{\chi }$ has the value indicated by Eq.~(\ref{eq:cond-up}),
for instance also at $u=1$. Using the asymptotic expansion of the
Bessel functions for large arguments, it is easy to prove that $\Omega
_{\chi }\propto x^{5/2}$ as claimed in the text.}. These
considerations justify the fact that the l.h.s. increases faster than
the r.h.s.; hence it suffices to evaluate the
inequality~(\ref{importantcc}) at the time of curvaton decay: if it is
then satisfied, it will have always been verified prior to
decay. After decay, of course, curvaton decay does not source anymore
the evolution of $\Omega_\chi$ and all perturbations stay constant, as
seen in previous figures. Since the time of curvaton decay is set by
$H\sim \Gamma_\sigma$, and $\Omega_\chi^{>_{\rm d}}\sim B_\chi
\Omega_\sigma^{<_{\rm d}}$, the previous condition of validity of the
analytical calculation can be simplified down to the same condition
(\ref{eq:cond-up}) than obtained in the sudden decay approximation. In
terms of our parameters, this reads:
\begin{equation}
B_\chi\Omega_\sigma^{\rm
(i)}\left(\frac{\Gamma_\sigma}{\Gamma_{\chi\chi \vert {\rm
f}}}\right)^{1/2}\,\ll\, 5.0\times10^{-3}x_{\rm f}^{1/2} {\rm
e}^{-x_{\rm f}}\ .\label{eq:cond3}
\end{equation}
Hence, the r.h.s. becomes $1.0\times10^{-11}$ for $x_{\rm f}=20$. For
the example considered above, namely
$\Gamma_\sigma/\Gamma_{\chi\chi\vert\rm f}=10^{-6}$ and
$B_\chi=10^{-1}$, condition (\ref{eq:cond3}) shows that the
isocurvature perturbation remains maximal as long as:
$\Omega_\sigma^{\rm (i)}\lesssim 4.6 \times 10^{-7}$, in good
agreement with Fig.~\ref{fig:f4a}.

\par
\begin{figure*}
  \centering
  \includegraphics[width=0.6\textwidth,clip=true]{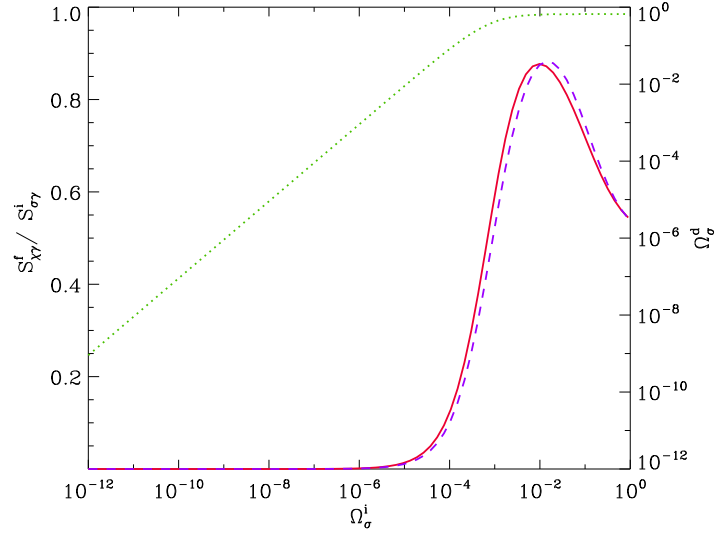}
  \caption[...]{Same as Fig.~\ref{fig:f4a}, but for $B_\chi=0$. }
\label{fig:f4b}
\end{figure*}

In order to model the effect of regenerated annihilations on the
perturbations, one can proceed as follows. One can study the evolution
of $\zeta _{\chi }$ and $\zeta _{\gamma }$ after curvaton decay,
assumed to be instantaneous for simplicity. In this case, one can
safely neglect the first term in Eq.~(\ref{background}) since it is
proportional to $\Omega _{\sigma }$ and the last term proportional to
$\Omega _{\chi ,{\rm eq}}$ as before. If one further considers that
radiation dominates (hence $\Omega _{\gamma }\simeq 1$), then
Eq.~(\ref{background}) can integrated exactly and the solution reads
\begin{equation}
\Omega _{\chi }=\Omega _{\chi }^{>_{\rm d}}\frac{\left(x/x_{\rm
d}\right)^2} {\Upsilon \left[\left(x/x_{\rm
d}\right)-1\right]+\left(x/x_{\rm d}\right)}\, ,
\end{equation}
where $x_{\rm d}$ is the time of curvaton decay.  Then, making the
same assumptions as before in Eq.~(\ref{evolzetachi}) and considering
that $\Phi $ remains constant (which is well-verified numerically),
one obtains
\begin{equation}
\Delta _{\chi }+\Phi=\left[\Delta _{\chi }^{>_{\rm d}}+\Phi^{_{\rm d}}\right]
\frac{\left(x/x_{\rm d}\right)} {\Upsilon \left[\left(x/x_{\rm
d}\right)-1\right]+\left(x/x_{\rm d}\right)}\, .
\end{equation}
Using this solution and the fact that $\Delta _{\gamma }\simeq -2\Phi
$ is constant, one can estimate the amplitude of the isocurvature
perturbations well after the curvaton decay. Straightforward
manipulations leads to
\begin{equation}
S_{\chi \gamma }^{\rm (f)}=\frac{1}{1+\Upsilon }S_{\chi \gamma
}^{>_{\rm d}} +\frac{\Upsilon }{2\left(\Upsilon +1\right)}\Phi^{>_{\rm
d}} \, .\label{eq:ann-S}
\end{equation}
This result indicates that the final isocurvature perturbation is
suppressed by a factor $(1+\Upsilon)$ which becomes all the more
important as annihilations are effective at curvaton decay. This is
reminiscent of the Weinberg theorem of the approach to adiabaticity
(\ie the erasure of isocurvature perturbations) of interacting fluids
(here, dark matter and radiation through annihilations), although the
physical conditions are here quite different, see
Appendix~\ref{app:adiabat}.

\begin{figure*}
  \centering
  \includegraphics[width=0.6\textwidth,clip=true]{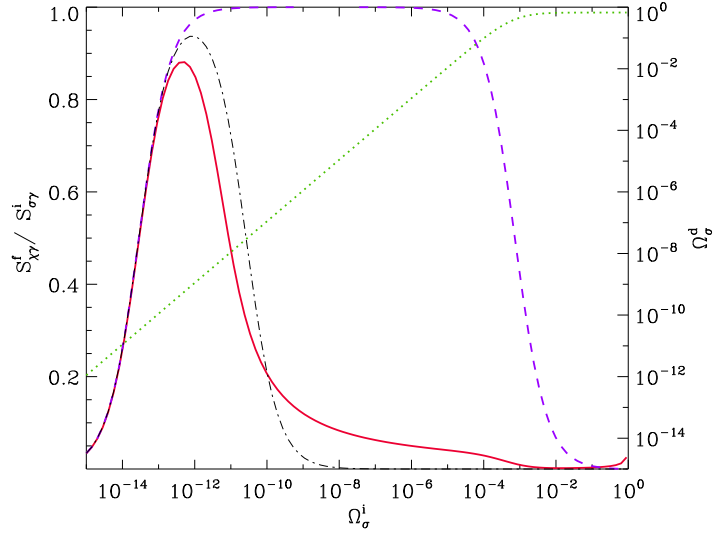}
  \caption[...]{Same as Fig.~\ref{fig:f4a}, but for $x_{\rm f}=30$.}
\label{fig:f4c}
\end{figure*}

\par

From our numerical experiments for different values of the parameters,
we find that this reduction factor  reproduces well the
numerical results.  For instance, in Fig.~\ref{fig:f4a}, the
suppression factor manifests itself by a reduction of $S_{\chi \gamma
}^{\rm (f)}$ for $\Omega_\sigma^{\rm (i)}\gta 4.6 \times 10^{-7}$, as
indicated by the dotted-dashed curve which agrees well with the
numerical calculations. However, the presence or functional form of
the second term in the r.h.s. of Eq.~(\ref{eq:ann-S}) is not confirmed
by numerical experiments in which $S_{\chi \gamma }^{>_{\rm d}}=0$;
one finds, in particular, $S_{\chi\gamma}^{\rm (f)}\simeq 0$ in this
case. This error is attributable to the assumption of instantaneous
curvaton decay, as we have checked numerically. Nevertheless, even if
this term is included in the formula (as done in the comparison to the
numerical results in Figs.~\ref{fig:f4a},\ref{fig:f4c}), the agreement
remains quite reasonable.

Let us now describe Fig.~\ref{fig:f4b}. It shows an altogether
different scenario, in which the branching ratio $B_\chi=0$, \ie the
curvaton decays only into radiation. In this case, significant
isocurvature is produced when $\Omega_\sigma^{<_{\rm d}}\sim 1$, as
discussed before. In this case, the analytical calculation always
agrees quite well with the numerical results. The isocurvature
perturbation tends to decrease as $\Omega_\sigma^{\rm (i)}$ exceeds
$\simeq 10^{-2}$ because the curvaton energy density at the time of
freeze-out [$\Omega_\sigma^{>_{\rm f}}\sim 6 \Omega_\sigma^{\rm (i)}$
for $x_{\rm f}=20$] then becomes significant. As a consequence, the
dark matter curvature perturbation is affected by that of the
curvaton, hence the final dark matter - radiation isocurvature is
reduced.

Let us turn to Fig.~\ref{fig:f4c}. It shows an example similar to
Fig.~\ref{fig:f4a} for a different value of $x_{\rm f}$, here $x_{\rm
f}=30$. The phenomenology is similar to that discussed for
Fig.~\ref{fig:f4a}; the main difference lies in the much lower values
of $\Omega_\sigma^{\rm (i)}$ where the isocurvature perturbation is
maximal. Since $x_{\rm f}$ is larger than in Fig.~\ref{fig:f4a}, the
post-freeze-out dark matter abundance is (much) smaller, hence a
lesser degree of contamination by curvaton decay is sufficient to
produce a substantial isocurvature perturbation. One can use the
formulae (\ref{eq:cond1}), (\ref{eq:cond2}) and (\ref{eq:cond3}) to
give a detailed account of Fig.~\ref{fig:f4c}.

\par

\begin{figure*}
  \centering
  \begin{tabular}{cc}
  \includegraphics[width=0.5\textwidth,clip=true]{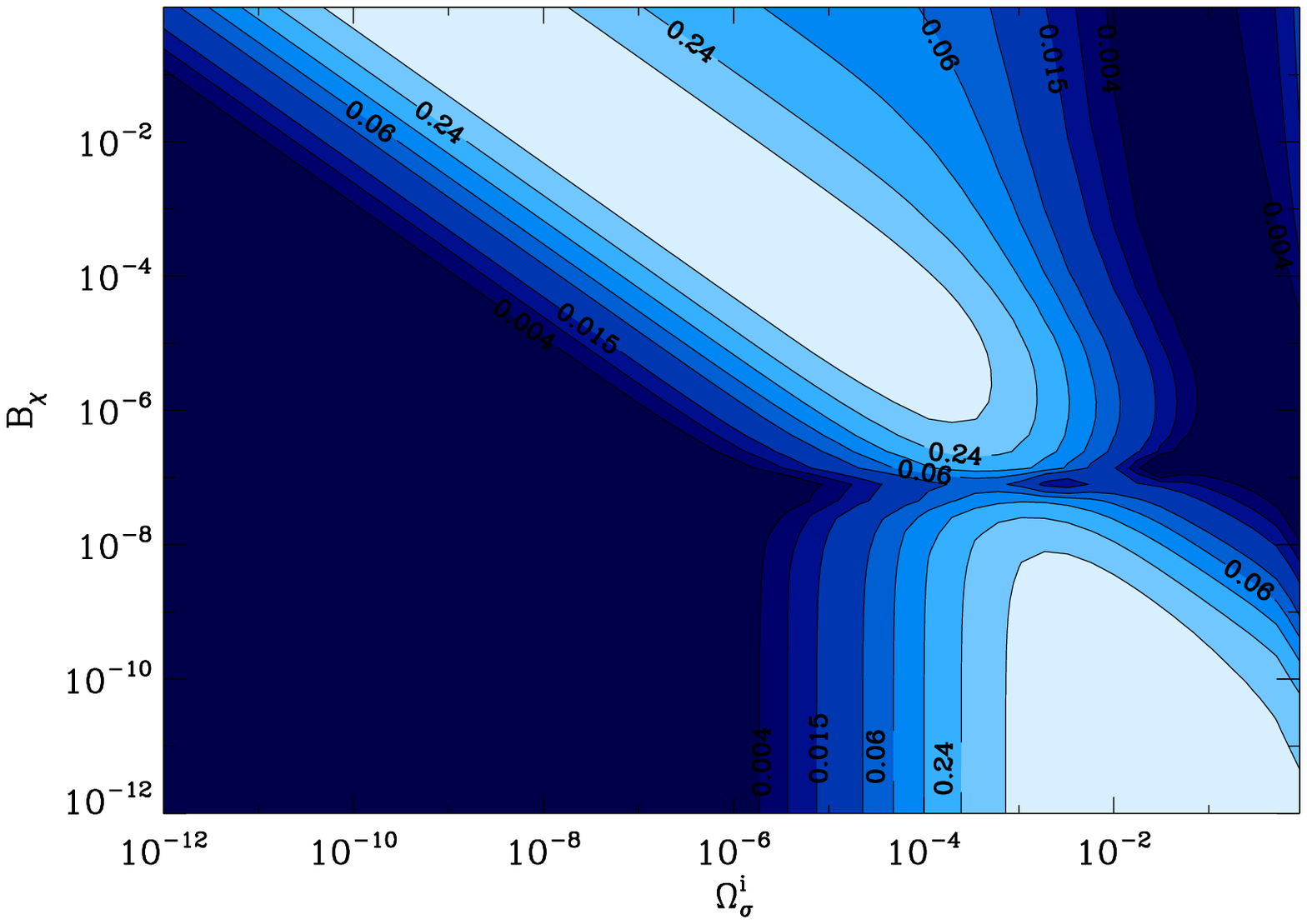} &
  \includegraphics[width=0.5\textwidth,clip=true]{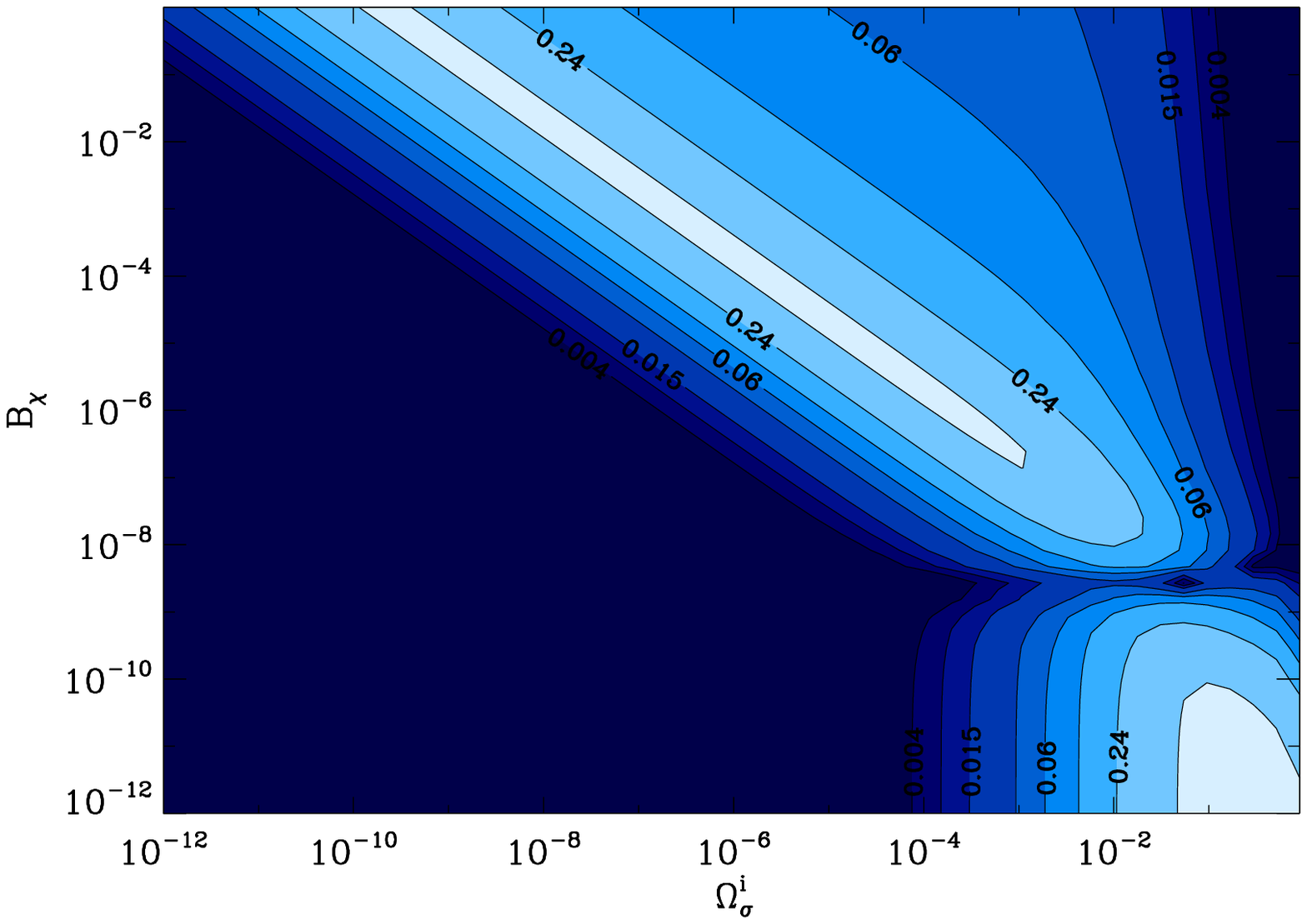}
  \end{tabular}
  \caption[...]{Contour plot of
  $S_{\chi\gamma}^{\rm(f)}/S_{\sigma\gamma}^{\rm(i)}$ in the plane
  $\Omega_\sigma^{\rm (i)}$, $B_\chi$, for $x_{\rm f}=20$. Each
  contour represents a decrement by a factor 2. In the left panel,
  $\Gamma_\sigma/\Gamma_{\chi\chi\vert\rm f}=10^{-6}$; in the right
  panel, $\Gamma_{\sigma}/\Gamma_{\chi\chi\vert\rm f}=10^{-3}$. }
\label{fig:f5a}
\end{figure*}

Finally, in a third series of figures, we show the same value
$S_{\chi\gamma}^{\rm(f)}/S_{\sigma\gamma}^{\rm(i)}$ in the plane of
the branching ratio $B_\chi$ and initial curvaton abundance
$\Omega_\sigma^{\rm (i)}$. The location of the peak of the
isocurvature abundance, as shown in Fig.~\ref{fig:f5a}, follows the
trend described above. Indeed, one sees two main regions where the
relative isocurvature perturbation is of order unity, one at large
branching ratio and small curvaton abundance, which corresponds to the
case in which the curvaton decay produces a substantial amount of dark
matter but negligible radiation, and the other at small branching
ratio but large curvaton abundance, corresponding to the opposite
limit.

The right panel of Fig.~\ref{fig:f5a} shows a plot similar to that in
the left panel, albeit for a much larger value
$\Gamma_{\sigma}/\Gamma_{\chi\chi\vert\rm f}=10^{-3}$, meaning that curvaton
decay occurs shortly after dark matter freeze-out (at $x_{\rm d}\sim
10^2$). Consequently, secondary dark matter annihilations due to
curvaton decay are more effective and restrain the part of parameter
space where the isocurvature perturbation is maximal. The lower region
is also pushed to larger values of $\Omega_\sigma^{\rm (i)}$ since, at
given $\Omega_\sigma^{\rm (i)}$, an earlier curvaton decay means a
smaller value of $\Omega_\sigma^{<_{\rm d}}$.


\section{Conclusion}
\label{Conclusion}

In this article, we have constructed a comprehensive model of curvaton
cosmology including radiation, WIMP dark matter (neutralinos) and a
massive curvaton field modeled as a perfect fluid with a dust-like
equation of state. We have computed the evolution of cosmological
fluctuations through dark matter freeze-out and curvaton decay
by approximate analytical methods as well as with a fully exact numerical
treatment. As a result, we determine the final curvature and
isocurvature perturbations and define the parameters that control
their amplitude.

\par

First of all, we have confirmed some results that are intuitively
clear. There is obviously no production of isocurvature perturbations
if the curvaton contribution is completely negligible or if, on the
contrary, radiation and dark matter both come entirely from curvaton
decay. Another case where the final $S_{\chi \gamma }$ vanishes is
when the curvaton decays before dark matter freeze-out: because dark
matter and radiation then share thermal equilibrium after curvaton
decay, the isocurvature mode is erased in agreement with the theorem
of Weinberg on the approach to adiabaticity for interacting
fluids~\cite{W04}.

\par

Our main finding is that a substantial radiation-dark matter
isocurvature mode can be produced if either the dark matter or the
radiation (but not both at the same time) mostly originates from
curvaton decay. Quantitatively, the first situation corresponds to
$B_{\chi }\Omega _{\sigma }^{<_{\rm d}}\gta \Omega _{\chi }^{<_{\rm
d}}$ and $\left(1-B_{\chi }\right)\Omega _{\sigma }^{<_{\rm d}}\ll
\Omega _{\gamma }^{<_{\rm d}}$ (we recall that $B_\chi$ denotes the
branching ratio of curvaton decay into dark matter, and that the
symbol $<_{\rm d}$ is intended to mean ``immediately before curvaton
decay''). Despite the fact that the curvaton is underdominant at its
decay, its influence on dark matter is important because the dark
matter energy density is itself even smaller. The second regime
corresponds to $B_{\chi }\sim 0$ and $\Omega _{\sigma }^{<_{\rm
d}}\gta \Omega _{\gamma }^{<_{\rm d}}$. In this case, at decay,
the curvaton perturbations are mainly transferred to radiation. The
curvaton perturbations also modifies the freeze-out at the
perturbative level, thereby contaminating the dark matter
fluctuations. Nevertheless, a significant radiation-dark matter
isocurvature mode (as measured relatively to the curvature
perturbation) is also produced in this limit.

\par

Another important effect that we have observed and modeled is that
regenerated dark matter at curvaton decay may annihilate, leading to a
partial (or even complete) erasure of any previously existing $S_{\chi
\gamma }$ isocurvature perturbation, even if the curvaton energy
density at that time is small, $\Omega _{\sigma }^{<_{\rm d}} \ll
1$. Assuming the curvaton decay to be instantaneous, we have provided
a new analytical formula describing this phenomenon which reproduces
fairly well our exact numerical estimates. These late time
annihilations further restrict the region in the parameter space of
the model (in particular the value of the initial curvaton energy
density) where significant isocurvature perturbations can be produced.

\appendix

\section{Perturbations to first order}
\label{app:perturb}

In this first appendix, one briefly recalls how the gauge-invariant
equations of motion of the (scalar) perturbations used in the main
text, see Eqs.~(\ref{pertchi})--(\ref{pertphi}), can be obtained. The
scalar part of the perturbed metric can be expressed as~\cite{KS84,
MFB92}
\begin{equation}
\label{ds2}
{\rm d}s^2 = a^2(\eta )\left \{- (1+2\phi){\rm d}\eta ^2 + 2\partial
_iB {\rm d}x^i {\rm d}\eta + \left[\left(1-2\psi\right)\delta
_{ij}+2\partial_i\partial _jE\right]{\rm d}x^i{\rm d}x^j\right \} \ ,
\end{equation}
and depends on four unknown functions: $\phi $, $B$, $\psi $ and
$E$. In the previous expression, $a(\eta )$ is the Friedmann-Lema\^
{\i}tre-Robertson-Walker scale factor and $\eta $ denotes the
conformal time. The most general (infinitesimal) change of coordinates
(or ``gauge-transformation'') that can be constructed with scalar
functions (given here by $\xi ^0$ and $\xi $) is given by
\begin{equation}
\eta \rightarrow \bar{\eta }=\eta +\xi ^0(\eta ,x^k),\quad 
x^i \rightarrow \bar{x}^i=x^i+\delta ^{ij}\partial _j\xi (\eta ,x^k)\, .
\end{equation}
Calculating the Lie derivative along the four vector defined above, we
find that the four scalar functions used to construct the scalar
perturbed metric transform according to
\begin{equation}
\label{liescalar}
\bar{\phi}=\phi +\xi ^{0'}+\frac{a'}{a}\xi ^0, \quad 
\bar{B}=B-\xi ^0+\xi ', \quad \bar{\psi }=\psi -\frac{a'}{a}\xi ^0, 
\quad \bar{E}=E+\xi \, ,
\end{equation}
where a prime denotes derivative with respect to conformal time. As a
consequence, the following combinations, known as the Bardeen
potentials
\begin{equation}
\Phi \equiv \phi +\frac{1}{a}\left[a\left(B-E'\right)\right]',\quad
\Psi \equiv \psi -\frac{a'}{a}\left(B-E'\right)\, ,
\end{equation}
are gauge invariant, that is to say, $\bar{\Phi }=\Phi$ and $\bar{\Psi
}=\Psi $. In particular, $\Phi $ can be viewed as the relativistic
generalization of the Newtonian potential. 

\par

One must also construct gauge-invariant combinations for the scalar
quantities appearing in the perturbed stress-energy tensor. The rule
of transformation of the stress-energy tensor is that of any two-rank
tensor and has already been given above. In the following, we consider
the case where several fluids are present and we denote each species
by the index ``$(\alpha )$''. Then, straightforward manipulations lead
to:
\begin{equation}
\bar{\delta \rho_{(\alpha)}}=\delta \rho_{(\alpha )}+\rho _{(\alpha)}'
\xi ^0\, \quad \bar{v_{(\alpha)}}=v_{(\alpha)}-\xi '\, \quad \bar{\delta
p_{(\alpha)}}=\delta p_{(\alpha )}+p _{(\alpha )}'\xi ^0\, ,
\end{equation}
where $\rho _{(\alpha) }$, $p_{(\alpha) }$ and $v_{(\alpha) }$ are
respectively the background energy density, pressure and velocity of
the fluid ``$(\alpha )$''. For the quantities describing the matter,
these expressions play the same role as Eqs.~(\ref{liescalar}) for the
metric perturbations. From these expressions, one can easily construct
gauge-invariant density contrast, pressure and velocity perturbations,
namely
\begin{equation}
\Delta \rho_{(\alpha)}\,=\,\delta \rho_{(\alpha)}
+\rho_{(\alpha)}'\left(B-E'\right)\, ,\qquad \Delta
p_{(\alpha)}\,=\,\delta p_{(\alpha)}+p_{(\alpha)}'(B-E'), \qquad
V_{(\alpha)}=v_{(\alpha)}+E'\, .
\end{equation}
The equations of motion for these quantities can be obtained from the
perturbed Einstein equations or from the time and space components
of the conservation equation, the two methods being equivalent thanks
to the Bianchi identities.

\par

In the presence of multiple interacting fluids, which is the case of
interest in the present article, one must include the various energy
transfer rates between the different fluids. One may express these
transfers in the form:
\begin{equation}
\nabla _{\mu}T^{\mu \nu}_{(\alpha )}=Q^{\nu }_{(\alpha
)}+Y^{\nu}_{(\alpha )}\, ,
\end{equation}
where the vector $Q^{\nu }_{(\alpha )}$ represents the case of a decay
and must be linear in energy density while $Y^{\nu }_{(\alpha )}$
describes the case of annihilations and must thus be quadratic in
energy density. This suggests the following covariant writing for the
transfer related to decay
\begin{equation}
\label{defQ}
Q^{\mu }_{(\alpha )}=\Gamma _{\sigma(\alpha )}T^{\mu
\nu}_{\sigma}u_{\nu}=-\Gamma _{\sigma(\alpha )}\rho _{\sigma}u^{\mu }\, ,
\end{equation}
$\Gamma _{\sigma(\alpha )}$ being the decay rate of species $\sigma$
to species $(\alpha)$; the above equations holds for a source, not a
loss term. In addition, the previous expression is valid if the
stress-energy tensor is that of a perfect fluid. As far as
annihilation terms are concerned, the analog $Y^\mu_{(\alpha )}$ of
the transfer rate $Q^\mu_{(\alpha )}$ can be written as
\begin{equation}
\label{defY}
Y^{\mu}_{(\alpha )}=\frac{\left \langle \sigma_{\chi\chi} v\right\rangle
_{(\alpha )}}{m_{(\alpha )}} \left[T^{\mu}_{(\alpha )}{}_{\lambda
}T^{\lambda \beta }_{(\alpha )}-T^{\mu }_{(\alpha ), {\rm
eq}}{}_{\lambda }T^{\lambda \beta }_{(\alpha ),{\rm
eq}}\right]u_{\beta } =\frac{\left \langle \sigma_{\chi\chi} v\right\rangle
_{(\alpha )} }{m_{(\alpha )}}\left[\rho ^2_{(\alpha )}-\rho _{(\alpha
),{\rm eq}}^2\right]u^{\mu }\, .
\end{equation}
In this expression $T^{\mu \nu}_{(\alpha ),{\rm eq}}$ is the
stress-energy tensor of the fluid ``$(\alpha )$'' in thermal
equilibrium. In particular, the corresponding energy density is given
by Eq.~(\ref{defneq}). As before, the last equation has been obtained
postulating a perfect fluid. Of course the total stress-energy tensor
is conserved and, therefore, one has:
\begin{equation}
\nabla _{\mu }\left[\sum _{\alpha }T_{(\alpha )}^{\mu \nu}\right]=\sum
_{\alpha }\left[Y_{(\alpha )}^{\nu }+Q_{(\alpha )}^{\nu}\right]=0\, .
\end{equation}

\par

One advantage of the previous covariant formalism is that the
transformation properties of the vectors $Q_{\mu }$ and $Y_{\mu }$ can
now be evaluated quite easily. Using Eqs.~(\ref{defQ})
and~(\ref{defY}), one finds:
\begin{eqnarray}
\delta Q^{\mu }_{(\alpha )} &=& \Gamma _{\sigma(\alpha )}\left[\delta
T^{\mu \nu}_{\sigma}u_{\nu }+ T^{\mu \nu}_{\sigma}\delta u_{\nu
}\right]\, , \\ \delta Y^{\mu}_{(\alpha )} &=& \frac{\langle \sigma_{\chi\chi}
v\rangle _{(\alpha )} }{m_{(\alpha )}} \Biggl[\delta T^{\mu }_{(\alpha
)}{}_{\lambda }T_{(\alpha )}^{\lambda \beta }u_{\beta } +T^{\mu
}_{(\alpha )}{}_{\lambda }\delta T_{(\alpha )}^{\lambda \beta
}u_{\beta } +T^{\mu }_{(\alpha )}{}_{\lambda }T_{(\alpha )}^{\lambda
\beta }\delta u_{\beta } -\delta T^{\mu }_{(\alpha ),{\rm
eq}}{}_{\lambda }T_{(\alpha ),{\rm eq}}^{\lambda \beta }u_{\beta }
-T^{\mu }_{(\alpha ),{\rm eq}}{}_{\lambda }\delta T_{(\alpha ),{\rm
eq}}^{\lambda \beta }u_{\beta } \nonumber \\& & -T^{\mu }_{(\alpha
),{\rm eq}}{}_{\lambda }T_{(\alpha ),{\rm eq}}^{\lambda \beta }\delta
u_{\beta }\Biggr] \, .
\end{eqnarray}
Working out the above expressions for the time component, this gives 
\begin{equation}
\label{timetransfer}
\delta Q_{(\alpha )0}=a\Gamma _{\sigma(\alpha )}\left[\delta \rho
_{\sigma}+\rho _{\sigma}\phi \right]\, , \qquad \delta Y_{(\alpha
)0}=\frac{\langle \sigma_{\chi\chi} v\rangle _{(\alpha )}}{m_{(\alpha
)}}\left\{-2a \left[\rho _{(\alpha )}\delta \rho_{(\alpha )} -\rho
_{(\alpha ),{\rm eq}}\delta \rho_{(\alpha ),{\rm
eq}}\right]-a\left[\rho ^2_{(\alpha )}-\rho ^2_{(\alpha ),{\rm
eq}}\right]\phi \right\}\, ,
\end{equation}
where we have used that $\delta u^0=-\phi/a$. Then, the time component
of the conservation equation leads to the following gauge-invariant
equation, for instance for the dark matter component:
\begin{eqnarray}
\label{timepertcons}
& & \Delta \rho_\chi'+3{\cal H} \left(\Delta \rho_\chi+\Delta
p_\chi\right) -3\left(\rho_\chi +p_\chi\right)\Phi' +\left(\rho_\chi
+p_\chi\right)\partial _k\partial ^kV_{\chi} = a\Gamma
_{\chi}\left(\Delta \rho_\chi +\rho_\chi \Phi\right) \nonumber \\ & &
-\frac{\langle \sigma_{\chi\chi} v\rangle }{m_\chi}a\left[ 2\rho_\chi
\Delta \rho_\chi-2\rho _{\chi,{\rm eq}}\Delta \rho_{\chi ,{\rm
eq}}+\left(\rho_\chi^2-\rho^2_{\chi ,{\rm eq}}\right)\Phi\right]\, .
\end{eqnarray}
On large scales the gradient of the velocity term can be
neglected. Then, the above equation exactly reduces to
Eq.~(\ref{pertchi}) used in the main body of the text. Other
equations~(\ref{pertgam}) and~(\ref{pertsig}) are obtained along the
same lines.

\section{Junction conditions}
\label{app:junct}

The junction conditions allow us to derive how the ``conserved''
quantities cross freeze-out. Assuming that the spatial transition
hypersurface $\Sigma$ is defined by the condition that some function
$q=q_0+\delta q$ remains constant, its normal $n^{\mu }$ is given by
$n_{\mu }=\partial _{\mu}q$ and, then, two conditions must be
fulfilled\cite{DM95,MS97} so that the two space-time manifolds along
$\Sigma$ can be joined without a surface layer: the induced spatial
metric $\perp_{ij}$ obtained from $\perp_{\mu \nu}\equiv g_{\mu \nu} +
n_{\mu } n_{\nu }$ and the extrinsic curvature (or second fundamental
form) $K_{ij}$ should be continuous on $\Sigma$, \ie
\begin{equation}
\left[\perp_{ij}\right]_{\pm}=0\, ,\quad \left[K_{ij}\right]_{\pm}=0\, .
\end{equation}
The extrinsic curvature is defined as:
\begin{equation}
\label{defK}
K_{\mu \nu}=-\frac{1}{2}{\cal L}_n \perp_{\mu \nu}
=-\frac{1}{2}\left(n^{\rho}\nabla _{\rho}\perp_{\mu \nu} +\perp_{\rho
\nu}\nabla _{\mu }n^{\rho } +\perp_{\rho \mu}\nabla _{\nu }n^{\rho
}\right)\, ,
\end{equation}
where ${\cal L}_n$ denotes the Lie derivative with respect to the
normal $n^{\mu}$. In order to compute $K_{ij}$ the system of
coordinates (\ie the gauge) and the vector $n^{\mu }$ (\ie the surface
of transition) have to be specified. Different choices for $n^{\mu}$
lead to inequivalent junction conditions. The previous equations are
the general rules that must applied in order to find the matching
conditions. Expressing the condition $\left[\delta\perp
_{ij}\right]_{\pm }=0$ for diagonal and off-diagonal terms leads to:
\begin{equation}
\label{psiE1}
\left[\psi\left(\eta_{_{\Sigma }}\right)\right]_\pm =\left[\psi +
{\cal H} \frac{\delta q}{q_0'}\right]_\pm\left(\eta _0\right)= 0\, ,
\qquad \left[E\left(\eta_{_{\Sigma }}\right)\right]_\pm
=\left[E\left(\eta _0\right)\right]_\pm =0\, ,
\end{equation}
where we have used that $\eta _{_{\Sigma }}=\eta _0+\delta \eta
$. Repeating the same calculation for the second fundamental form
gives two other (gauge-invariant) junction conditions, namely
\begin{equation}
\label{JCK}
\left[\psi '+{\cal H}\phi +\left({\cal H}'-{\cal
H}^2\right)\frac{\delta q}{q_0'}\right]_\pm\left(\eta _0\right)=0\, ,
\qquad \left[B-E'+\frac{\delta q}{q_0'}\right]_\pm\left(\eta
_0\right)=0\, .
\end{equation}
The equations (\ref{psiE1}) and (\ref{JCK}) represent the complete set
of matching conditions for density perturbations. Let us now derive
the consequences of these equations. Since $\left[\psi
\right]_{\pm}=-{\cal H}\left[\delta q/q_0'\right]_{\pm }$,
$\left[B-E'\right]_{\pm }=-\left[\delta q/q_0'\right]_{\pm }$ and
$\left[\Psi \right]_{\pm }=\left[\Phi \right]_{\pm } =\left[\psi
\right]_{\pm }-{\cal H}\left[B-E'\right]_{\pm }$ (the first equation
of this last relation simply comes from the fact that we assume $\Phi
=\Psi$), we deduce that
\begin{equation}
\left[\Phi \right]_{\pm }=0\, .
\end{equation}
This condition does not depend on which quantity $q$ we choose and is
therefore hypersurface independent. On the other hand, to obtain the
other condition, we must specify this quantity. For the particular
case of dark matter annihilation freeze-out, the transition is
determined by the quantity
\begin{equation}
q_0=\frac{\Gamma }{H} \, .
\end{equation}
Before freeze out one has $\Gamma /H\propto T^{\alpha }/\rho ^{1/2}$
and this leads to
\begin{equation}
\left(-\psi -{\cal H}\frac{\delta q}{q_0'}\right \vert _{<_{\rm f}}
=\zeta _{\gamma }^{<_{\rm f}}+\frac{3\Omega _{\sigma }\left(\zeta
_{\sigma }^{<_{\rm f}} -\zeta _{\gamma }^{<_{\rm
f}}\right)}{\left[2\left (2-\alpha \right)-\Omega _{\sigma }
\right]}+{\cal O}\left(\Omega _{\chi }\right)\, .
\end{equation}
After freeze-out, one has $\Gamma n_{\chi }\langle \sigma_{\chi\chi}
v\rangle \propto \rho _{\chi }$. This leads to
\begin{equation}
\left(-\psi -{\cal H}\frac{\delta q}{q_0'}\right \vert _{>_{\rm f}} =
\zeta _{\chi }^{>_{\rm f}}+\frac{2}{2+\Omega _{\sigma }}
\left[\left(2-\frac12\Omega _{\sigma }\right)\zeta _{\chi }^{>_{\rm
f}} -\frac32\Omega _{\sigma }\zeta _{\sigma }^{>_{\rm
f}}-2\left(1-\Omega _{\sigma }\right) \zeta _{\gamma }^{>_{\rm
f}}\right]\, .
\end{equation}
Equating these two expressions and using the fact that $\zeta _{\gamma
}^{<_{\rm f}} =\zeta _{\gamma }^{>_{\rm f}}$ and $\zeta _{\sigma
}^{<_{\rm f}} =\zeta _{\sigma }^{>_{\rm f}}$, one arrives at
\begin{equation}
\zeta _{\chi }^{>_{\rm f}}=\zeta _{\gamma}^{<_{\rm
f}}+\frac{\left(\alpha -3\right) \Omega _{\sigma \vert {\rm
f}}}{2\left(\alpha -2\right)+\Omega _{\sigma \vert {\rm f}}}
\left(\zeta _{\sigma }^{<_{\rm f}}-\zeta _{\gamma }^{<_{\rm
f}}\right)\, ,
\end{equation}
where the background quantities are evaluated at the time of
freeze-out and where we recall that $\alpha =m/T+3/2$ evaluated at the
freeze-out. We have thus recovered, by means of the junction
conditions, Eq.~(\ref{eq:LW}) used in the text and derived for the
first time (although using different techniques) in Ref.~\cite{LW03}.

\section{Approach to adiabaticity for interacting fluids}
\label{app:adiabat}

In this last appendix, we discuss the theorem proved by
Weinberg~\cite{W04}, according to which any pre-existing isocurvature
perturbation between two interacting fluids is rapidly erased at
thermal equilibrium. In addition, we explicitly calculate the typical
time scale of this phenomenon in the present context and compare the
analytical prediction to numerical calculations.

\par

The number density of dark matter particles evolves according to
$n'+3{\cal H}n=Y_0/m_{\chi}$, with $Y_0/m_{\chi }=-a\langle
\sigma_{\chi\chi} v\rangle \left(n^2-n_{\rm eq}^2\right)$, in the
absence of curvaton decays but accounting for $\chi-\chi$
annihilations. This equation cannot be solved explicitly but useful
information can be obtained by linearizing it. For this purpose, one
writes $n=n_{\rm eq}+\delta n$ where $\delta n\ll n_{\rm eq}$. Then,
the equation obeyed by $\delta n$ reads, to first order in $\delta
n/n_{\rm eq}$:
\begin{equation}
\frac{{\rm d}}{{\rm d}t}\left(a^3\delta n\right)+2\langle
\sigma_{\chi\chi} v\rangle n_{\rm eq}a^3 \delta n=-\frac{{\rm d}}{{\rm
d}t}\left(a^3n_{\rm eq}\right)\, ,
\end{equation}
whose solution can be expressed as:
\begin{equation}
\delta n\,=\, \delta n(t_{\rm i})\left[\frac{a(t_{\rm
    i})}{a(t)}\right]^3\eta(t_{\rm i};\,t) - \frac{1}{a^3}\int_{t_{\rm i}}^t{\rm
    d}\theta\frac{\rm d}{{\rm d}\theta}\left(a^3n_{\rm
    eq}\right)\eta(\theta ;\, t)\ ,\label{eq:deltan}
\end{equation}
where we have defined the function:
\begin{equation}
\eta(t' ;\, t)\,\equiv\,\exp\left(-2\int_{t'}^t{\rm d}\tau\,
  \langle\sigma_{\chi\chi}v\rangle n_{\rm eq}\right)\ .
\end{equation}
As long as $x<x_{\rm f}$, the integrand (annihilation rate)
$\langle\sigma_{\chi\chi}v\rangle n_{\rm eq}$ is very large compared
to an inverse Hubble time, hence the function $\eta(t';\, t)$
collapses to zero on a very short timescale $\sim
\left[\langle\sigma_{\chi\chi}v\rangle n_{\rm eq}\right]^{-1}$. One
can thus Taylor expand the integral to first order and write:
\begin{equation}
\eta(t' ;\,
t)\,\simeq\,\exp\left[-(t-t')\times2\langle\sigma_{\chi\chi}v\rangle
n_{\rm eq}\right]\,\approx\, \frac{1}{\langle \sigma_{\chi\chi}
v\rangle n_{\rm eq}(t')}\delta\left(t-t'
\right)\,\label{eq:approxdelta}
\end{equation}
using the following representation of the Dirac function, $\delta
_{\epsilon}(x)={\rm e}^{-\vert x\vert/\epsilon}/(2\epsilon )$.

\par

The first term on the r.h.s. of Eq.~(\ref{eq:deltan}) corresponds to
the difference at $t_{\rm i}$ between $n_{\rm eq}(t_{\rm i})$ and the
solution of the evolution equation for $n$. This term collapses on the
above short timescale due to the presence of the function $\eta(t_{\rm
i}; t)$. This timescale can be written in a compact way in terms of
the $e-$fold number $N$, using the change of variables ${\rm d}t\,=\,
{\rm d}N/H$:
\begin{equation}
\eta(t_{\rm i} ;\, t)\,\simeq\,\exp\left(-\frac{N-N_{\rm i}}{\Delta
    N}\right)\ ,\label{eq:etafunc}
\end{equation}
with:
\begin{equation}
\Delta N\,=\,\frac{H(t_{\rm
    i})}{2\langle\sigma_{\chi\chi}v\rangle_{t_{\rm i}} n_{\rm
    eq}(t_{\rm i})}\,\simeq \,\frac{1}{2\sqrt{x_{\rm f}x_{\rm i}}}{\rm
    e}^{-x_{\rm f}+x_{\rm i}}\ .\label{eq:defold}
\end{equation}
For $x_{\rm i}=3$ and $x_{\rm f}=20$, the number of $e-$folds
necessary to erase any initial departure from the solution $n$ is:
$\Delta N\simeq 2.67\times10^{-9}$, very small indeed as compared to
unity.

The second term on the r.h.s. of Eq.~(\ref{eq:deltan}) represents the
difference between $n$ and $n_{\rm eq}$. It will prove interesting for
the discussion that follows to estimate the relative magnitude of this
term. Assuming that $\delta n_{\rm i}=0$, one can re-express this term
by changing variables ${\rm d}t={\rm d}\ln(x)/H$ in
Eq.~(\ref{eq:deltan}):
\begin{equation}
\delta n(x)\,=\, \int_{x_{\rm i}}^{x}{\rm d}x'
\left(\frac{x'}{x}\right)^3\frac{x'-3/2}{x'}n_{\rm
eq}(x')\eta(x';\,x)\,\simeq\,
\frac{1}{2}\frac{x-3/2}{x}\left(\frac{x}{x_{\rm f}}\right)^{1/2}{\rm
e}^{x-x_{\rm f}}n_{\rm eq}(x)\ ,
\end{equation}
where we have approximated $\eta(x';\,x)$ with the delta function in
the last equality (the factor of $1/2$ comes from the fact that the
Dirac function is peaked over the upper boundary of the
integral). Hence $\delta n/n_{\rm eq}\sim {\cal O}\left(e^{x-x_{\rm
f}}\right)\ll 1$ when $x< x_{\rm f}$.

Finally, by integrating by parts Eq.~(\ref{eq:deltan}), one can write
the global solution $n$ as:
\begin{equation}
\label{solin}
n(t)\simeq \frac{n(t_{\rm i})}{a^3}\eta(t_{\rm i};\,t)
 +\frac{2}{a^3}\langle \sigma_{\chi\chi}
v\rangle\int _{t_{\rm i}}^t a^3\left(\theta \right) n_{\rm
eq}^2\left(\theta \right)\eta(\theta;\,t)\,{\rm d}\theta .
\end{equation}
Here as well, the first term on the r.h.s. represents the initial
value which collapses on a time scale
$(2\langle\sigma_{\chi\chi}v\rangle n_{\rm eq})^{-1}$, while the
second term can be shown to approximate $n_{\rm eq}$ to within the
above $\delta n$.

\par

Let us now turn the perturbed quantities. At the perturbed level the
equation $\nabla _{\mu }n^{\mu }=0$ gives
\begin{equation}
\label{nablan0}
\partial _0\delta n^0+\partial _i\delta n^i+4{\cal H}\delta n^0
+\left(-3\psi '+\partial _i\partial ^iE'+\phi'\right)n^0=0\, .
\end{equation}
The quadrivector $n_{\mu}$ can be written as $n_{\mu}=nu_{\mu }$ and,
therefore, one has $\delta n_0=-a\left(\delta n+n\phi \right)$. Then, 
one can define the following gauge-invariant quantity 
\begin{equation}
\Delta n=\delta n+n'\left(B-E'\right)\, ,
\end{equation}
and, on large scales, Eq.~(\ref{nablan0}) reads
\begin{equation}
\Delta n'+3{\cal H}\Delta n-3n\Psi '=0\, .
\end{equation}
This equation is nothing but Eq.~(\ref{timepertcons}) for a
pressureless fluid since then $\rho =m_{\chi }n$. Again if one takes into
account the interaction term, then the above equation becomes
\begin{equation}
\Delta n'+3{\cal H}\Delta n-3n\Phi '=\frac{\delta Y_0}{m_{\chi }}\, ,
\end{equation}
where (using again the relation $\rho=m_{\chi }n$) the explicit form
of the annihilation term is given by Eq.~(\ref{timetransfer}), namely
\begin{equation}
\frac{\delta Y_0}{m_{\chi }}=-2a\langle \sigma_{\chi\chi} v\rangle
\left(n \Delta n-n_{\rm eq}\Delta n_{\rm eq}\right)
+\frac{Y_0}{m_{\chi }}\Phi \, ,
\end{equation}
where $\Delta n_{\rm eq}=n_{\rm eq}\Delta_{\rm eq}$, and $\Delta_{\rm
eq}$ has been defined in Eq.~(\ref{defdneq}). In particular, since
$\Delta n_{\rm eq}$ is proportional to $\Delta \rho _{\gamma}$, its
gauge-invariance is automatically guaranteed. In Ref.~\cite{LW03}, the
following quantity
\begin{equation}
\tilde{\zeta}_{\chi }\equiv -\Phi -{\cal H}\frac{\Delta n}{n'}\, ,
\end{equation}
has been defined. This quantity is obviously gauge-invariant and, if
$Q_{\mu }$ and $Y_{\mu }$ are not considered, is simply equal to
$\zeta _{\chi }$. Then, let us consider the following combination
\begin{equation}
\frac{{\rm d}}{{\rm d}\eta }\left[ -\frac{\Delta \rho _{\gamma }}{\rho
    _{\gamma }'} \left(n'+3{\cal H}n-\frac{Y_0}{m_{\chi
  }}\right)\right]+ \Delta n'+3{\cal H}\Delta n-3n\Phi '-\frac{\delta
  Y_0}{m_{\chi }}=0 \, .
\end{equation}
Working out this expression, one obtains
\begin{eqnarray}
\label{exact}
\frac{{\rm d}}{{\rm d}\eta }\left(\Delta n-n'\frac{\Delta \rho
_{\gamma }}{\rho _{\gamma }'}\right) &=& -3{\cal H}\left(\Delta
n-n'\frac{\Delta \rho _{\gamma }}{\rho _{\gamma }'}\right) -2a\langle
\sigma_{\chi\chi} v\rangle n\Delta n +2a\langle \sigma_{\chi\chi}
v\rangle n_{\rm eq}\Delta n _{\rm eq}+\frac{Y_0}{m_{\chi
}}\left[\Phi-\left(\frac{\Delta \rho _{\gamma }}{\rho _{\gamma
}'}\right)'\right] \nonumber \\ & & -\frac{Y_0'}{m_{\chi
}}\frac{\Delta \rho _{\gamma }}{\rho _{\gamma }'}+3n\left[\Phi'+{\cal
H}'\frac{\Delta \rho _{\gamma }}{\rho _{\gamma }'} +{\cal
H}\left(\frac{\Delta \rho _{\gamma }}{\rho _{\gamma
}'}\right)'\right]\, .
\end{eqnarray}
This formula is exact. At this stage, one should make some
approximations adapted to the situation at hands. Let us first assume
that we are before freeze-out. According to the previous
considerations, one can neglect $Y_0$ since its magnitude is of order
$\delta n/n_{\rm eq}\ll1 $ ($x<x_{\rm f}$) as compared to the
others. The previous equation then reduces to
\begin{eqnarray}
\frac{{\rm d}}{{\rm d}\eta }\left(\Delta n-n'\frac{\Delta \rho
_{\gamma }}{\rho _{\gamma }'}\right) &=& \left(-3{\cal H}-2a\langle
\sigma_{\chi\chi} v\rangle n_{\rm eq}\right) \left(\Delta
n-n'\frac{\Delta \rho _{\gamma }}{\rho _{\gamma }'}\right)
+3n\left[\Phi'+{\cal H}'\frac{\Delta \rho _{\gamma }}{\rho _{\gamma
}'} +{\cal H}\left(\frac{\Delta \rho _{\gamma }}{\rho _{\gamma
}'}\right)'\right]\, .
\end{eqnarray}

\begin{figure*}
  \centering \includegraphics[width=0.6\textwidth,clip=true]{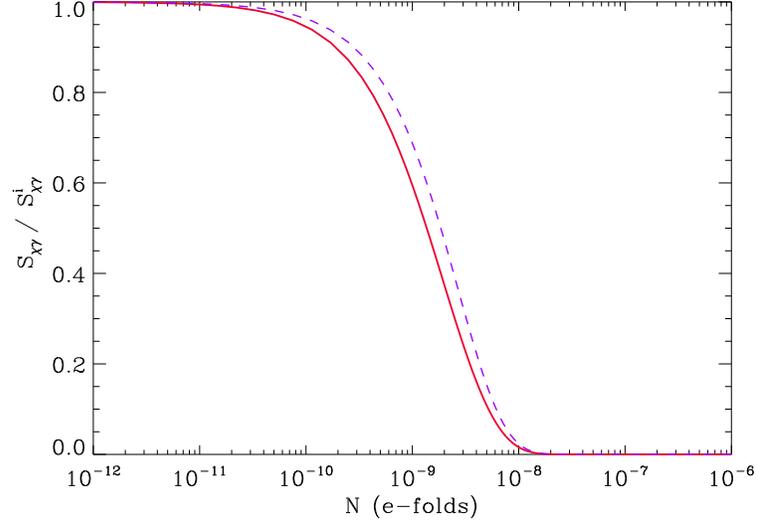}
  \caption[...]{Evolution of the isocurvature perturbation
  $S_{\chi\gamma}$ [relative to the initial $S_{\chi\gamma}^{\rm
  (i)}$] during the first fractions of $e-$folds at $x\simeq 3$. The
  thick solid line shows the results of the numerical integration, and
  the dashed line that of the analytical calculation, given by
  Eq.~(\ref{eq:theq}).  }
\label{fig:f1}
\end{figure*}

Now, we further assume that freeze-out occurs in a radiation-dominated
era. In this case, the quantity defined by $\zeta _{\gamma }\equiv
-\Phi -{\cal H}\Delta \rho _{\gamma }/\rho _{\gamma }'$ is conserved
hence the term proportional to $3n$ vanishes. Therefore, using the
fact that $n'\simeq-{\cal H}n\left(x+3/2\right)$, the
final equation reads
\begin{eqnarray}
\frac{{\rm d}}{{\rm d}\eta }\left[\ln \left(\frac{\Delta n
}{n}-\frac{n'}{n}\frac{\Delta \rho _{\gamma }}{\rho _{\gamma
}'}\right) \right] &=& \left(x-\frac32\right){\cal H}
-2a\langle \sigma_{\chi\chi} v\rangle n_{\rm
eq}\, .
\end{eqnarray}
However, before freeze-out, the first term in the right hand side of
the above relation is much smaller than the second one and, hence, can
be safely neglected. Finally, one arrives at
\begin{equation}
\tilde{\zeta }_{\chi }-\zeta _{\gamma }= \left(\tilde{\zeta }_{\chi
}-\zeta _{\gamma }\right)_{\rm i} \frac{x_{\rm i}+3/2}{x+3/2}
\eta(t_{\rm i};\,t)\, ,\label{eq:theq}
\end{equation}
where the function $\eta(t_{\rm i};\,t)$ has already been computed in
Eqs.~(\ref{eq:approxdelta}) and (\ref{eq:etafunc}).

This analytical result is compared to the numerical calculations in
Fig.~\ref{fig:f1}. Indeed, the agreement is excellent, and the
numerical calculations confirm the erasure of any initial isocurvature
perturbation $S_{\chi\gamma}$ on the $e-$fold scale $\Delta N\sim
3\times 10^{-9}$ computed in Eq.~(\ref{eq:defold}).



\begin{thebibliography}{99}
\bibitem{BLO} A curvaton scenario was suggested in Ref.~\cite{BLO00}
  as a way-out to overcome the difficulties of the pre-big-bang
  scenario in generating adiabatic perturbations; the corresponding
  calculations were provided in ~\cite{ES02}.
\end{thebibliography}
\end{document}